\PassOptionsToPackage{prologue,dvipsnames}{xcolor}
\documentclass[sigplan,10pt]{acmart}

\acmYear{2025}\copyrightyear{2025}
\setcopyright{rightsretained}
\acmConference[EuroSys '25]{Twentieth European Conference on Computer Systems}{March 30--April 3, 2025}{Rotterdam, Netherlands}
\acmBooktitle{Twentieth European Conference on Computer Systems (EuroSys '25), March 30--April 3, 2025, Rotterdam, Netherlands}
\acmDOI{10.1145/3689031.3717480}
\acmISBN{979-8-4007-1196-1/25/03}



\usepackage{color,soul}
\usepackage{hyperref}
\usepackage[normalem]{ulem}
\usepackage{enumitem}
\usepackage{subcaption}
\usepackage{amsmath}
\usepackage[noend]{algpseudocode}
\usepackage{algorithm}
\usepackage[]{mdframed}
\usepackage[utf8]{inputenc}   
\usepackage[T1]{fontenc}      
\usepackage{hyperref}         
\usepackage{url}              
\usepackage{booktabs}         
\usepackage{nicefrac}         
\usepackage{amsfonts}         
\usepackage{microtype}        
\usepackage{graphicx}         
\usepackage{sidecap, caption} 
\usepackage{multirow}         
\usepackage{enumitem}         
\usepackage{mathtools}
\usepackage{dsfont}
\usepackage{xspace}
\usepackage{tikz}
\usepackage{pifont}

\usepackage[toc,page]{appendix}
\newcommand{\eat}[1]{}

\newcommand{\changed}[1]{{#1}}
\newcommand{\added}[1]{{#1}}

\newcommand{\autotuning}{autotuning }
\newcommand{\Autotuning}{Autotuning }

\newcommand{\postgres}{PostgreSQL }
\newcommand{\redis}{Redis }
\newcommand{\nginx}{NGINX }

\newcommand{\join}{\texttt{JOIN} }
\newcommand{\joins}{\texttt{JOIN}s }

\newcommand{\Section}{{\S}}

\algrenewcommand\algorithmicrequire{\textbf{Input:}}
\algrenewcommand\algorithmicensure{\textbf{Output:}}
\algnewcommand\algorithmicforeach{\textbf{for each}}
\algdef{S}[FOR]{ForEach}[1]{\algorithmicforeach\ #1\ \algorithmicdo}

\newcommand{\cmark}{\color{ForestGreen}{\ding{51}}}%
\newcommand{\xmark}{\color{red}{\ding{55}}}%

\makeatletter
\newcommand\bigDiamond{\mathop{\mathpalette\bigDi@mond\relax}}
\newcommand\bigDi@mond[2]{%
  \vcenter{\hbox{\m@th
    \scalebox{\ifx#1\displaystyle 2\else1.2\fi}{$#1\Diamond$}%
  }}%
}
\newcommand\bigLozenge{\mathop{\mathpalette\bigL@zenge\relax}}
\newcommand\bigL@zenge[2]{%
  \vcenter{\hbox{\m@th
    \scalebox{\ifx#1\displaystyle 2\else1.2\fi}{$#1\blacklozenge$}%
  }}%
}

\newcommand{\TakeawayBox}[1]{
\vspace{.4em}
\noindent
\begin{minipage}{\linewidth}
\begin{mdframed}[linecolor=black,linewidth=1pt,innerleftmargin=8pt,innerrightmargin=8pt,innertopmargin=4pt,innerbottommargin=4pt, skipabove=8pt]
{\Large $\blacklozenge$ }
#1
\end{mdframed}
\end{minipage}
}

\newcommand{\SystemName}{TUNA\xspace}

\newcommand{\Dsv}{\texttt{D8s\_v5 }}

\begin{document}
\sloppy

\title{\SystemName: Tuning Unstable and Noisy Cloud Applications}

\author{Johannes Freischuetz}
\email{freischuetz@cs.wisc.edu}
\orcid{0009-0004-2667-4852}
\affiliation{%
  \institution{University of Wisconsin -- Madison}
  \city{Madison}
  \state{Wisconsin}
  \country{USA}
  \postcode{53706}
}

\author{Konstantinos Kanellis}
\email{kkanellis@cs.wisc.edu}
\orcid{0009-0006-3776-3863}
\affiliation{%
  \institution{University of Wisconsin -- Madison}
  \city{Madison}
  \state{Wisconsin}
  \country{USA}
  \postcode{53706}
}

\author{Brian Kroth}
\email{bpkroth@microsoft.com}
\orcid{0000-0002-5108-6743}
\affiliation{%
  \institution{Microsoft Gray Systems Lab}
  \city{Madison}
  \state{Wisconsin}
  \country{USA}
  \postcode{53703}
}

\author{Shivaram Venkataraman}
\email{shivaram@cs.wisc.edu}
\orcid{0000-0001-9575-7935}
\affiliation{%
  \institution{University of Wisconsin -- Madison}
  \city{Madison}
  \state{Wisconsin}
  \country{USA}
  \postcode{53706}
}

\begin{abstract}
\Autotuning plays a pivotal role in optimizing the performance of systems, particularly in large-scale cloud deployments.
One of the main challenges in performing \autotuning in the cloud arises from performance variability. 
We first investigate the extent to which noise slows \autotuning and find that as little as $5\%$ noise can lead to a $2.5$x slowdown in converging to the best-performing configuration.
We measure the magnitude of noise in cloud computing settings and find that while some components (CPU, disk) have almost no performance variability,
there are still sources of significant variability (caches, memory).
Furthermore, variability leads to \autotuning finding \emph{unstable} configurations.
As many as $63.3\%$ of the configurations selected as "best" during tuning can have their performance degrade by $30\%$ or more when deployed.
Using this as motivation, we propose a novel approach to improve the efficiency of \autotuning systems by (a) detecting and removing outlier configurations and (b) using ML-based approaches to provide a more stable \emph{true} signal of de-noised experiment results to the optimizer.
The resulting system, \SystemName (\underline{T}uning \underline{U}nstable and \underline{N}oisy Cloud \underline{A}pplications) enables faster convergence and robust configurations.
Tuning \postgres running \texttt{mssales}, an enterprise production workload, we find that \SystemName can lead to $1.88$x lower running time on average with $2.58x$ lower standard deviation compared to traditional sampling methodologies.
\end{abstract}

\maketitle

\setlength{\belowcaptionskip}{-5pt}

\section{Introduction}
\label{sec:introduction}
\begin{figure}[t]
    \centering
    \includegraphics[width=.9\linewidth,page=1]{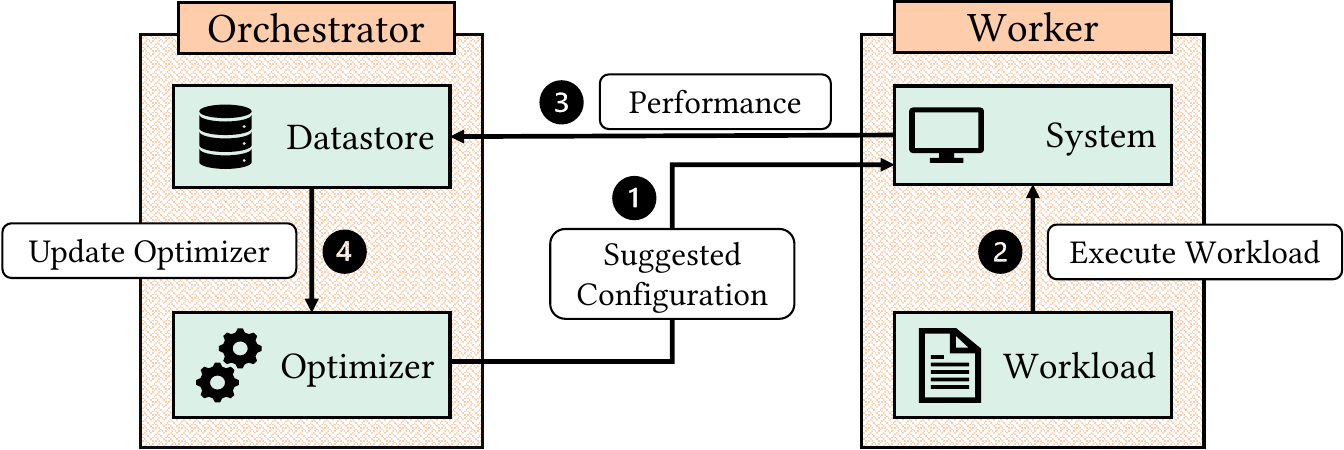}
    \caption{Modern auto-tuning system design~\cite{Llamatune, Ottertune, CDBTune}}
    \label{fig:traditionaldesign}
\end{figure}

Software tuning is important for many different types of systems.
Before the rise in popularity of machine learning (ML), heuristics and search-based approaches for tuning were used in different areas including databases~\cite{ituned, db2_tuning_no_ml, sql_server_tuning}, key-value stores~\cite{cosine_kv}, VM sizing~\cite{select_best_vm}, compilers~\cite{compiler_tuning}, distributed systems~\cite{hadoop_tuning_no_ml}, and memory systems\cite{memory_tuning_no_ml}.
With the rise in popularity of applying ML to systems problems, several recent works have developed \autotuning frameworks for databases (e.g.,  OtterTune~\cite{OtterTuneRealWorld}, CDBTune~\cite{CDBTune}, QTune~\cite{QTune}),
file systems~\cite{storagetuning} and analytics frameworks~\cite{MLScheduler}.

Most state-of-the-art tuners follow a similar design for \autotuning as shown in Figure~\ref{fig:traditionaldesign}.
The \autotuning process consists of a short loop:
first, an \emph{optimizer} (e.g., Bayesian or Gaussian Process optimizer etc.) suggests a new configuration (config) to evaluate for a System-under-Test (SuT) running a specific workload.
Typically, an initialization set, a set of randomly selected configurations, is used to bootstrap the model.
After the initialization set, a metric such as Expected Improvement or Thompson sampling~\cite{tohmpson_sampling} is used to generate the new suggestions~\cite{SMAC}.
Each suggested config is evaluated (sometimes called "profiling") for a fixed period to measure its performance.
The results are then cataloged, along with all prior evaluations, and then finally returned to the optimizer to improve future suggestions.
Once a given stopping criteria is met (e.g., total tuning time or number of configs), the best config is selected from the catalog.

Although ML-based solutions have been shown to achieve significant \emph{peak} performance improvements
(e.g., 2-5x~\cite{OtterTuneRealWorld} performance improvement in database systems, 20-25\%~\cite{storagetuning} performance improvement in storage systems, and 15-20\%~\cite{clustertuning} improvement in cluster scheduling), existing work has not studied if tuning produces configs with stable and predictable performance in the cloud (i.e., performance remains the same when deployed on similar hardware).
This conflicts with service level agreement (SLA)~\cite{define_sla} requirements, which demand stable performance on customer workloads.

We find that there are two main challenges in finding stable, well-performing configs: first from platform level \emph{performance variability} during tuning, and second from \emph{unstable behavior} of the SuT. 
Given that various levels of the hardware and software stack have 
been shown to have variable performance~\cite{TamingPerformanceVariability, GPUVariance, CpuCorePerfVariance, HPCVariability, WindowsServerDiskVariance, CloudPerfVariability}, when a config suggested by the optimizer is evaluated, the measured performance of the SuT can be noisy.
We find that even having a small amount of measurement noise (5\%) can lead to a significant slowdown (2.5$\times$) in finding good performing configs (\S\ref{sec:motivation}).
Additionally, we find that the software systems can exhibit unstable behavior with specific configurations; typically this can be attributed to configurations changing which code paths are taken at runtime, leading to unexpected performance.
We first expand on these findings by discussing our benchmark of cloud computing offerings to measure the amount of variability caused by hardware components, virtualization, and noisy neighbors and follow that with a case study on how unstable behavior can affect autotuning.

To measure variability in the cloud, we run a $68$ week-long study, sampling across over $43500$ VMs on Microsoft Azure.
While there have been prior cloud measurement studies (listed in Table~\ref{table:CloudStudy}), these studies are either too old~\cite{EC2PerfVariance, LongCloudPerfProd, Moreforyourmoney} and do not capture recent changes to the cloud, or do not sample a large set of VMs to understand the distribution of performance across a region~\cite{EC2PerfVariance, Moreforyourmoney, IaaSVariance, TamingPerformanceVariability, CloudBenchmarkSuite, bigdatareproducable, NoiseInTheClouds}.
To the best of our knowledge, this is the longest, and largest cloud study conducted and we find, contrary to prior work, very consistent CPU and disk performance.
The variability of these components using the Coefficient of Variation (CoV) (the standard deviation normalized by the mean), is less than $1.0\%$.
This is significantly lower than the results reported for bare-metal machines ($9.0\%$ for disk~\cite{TamingPerformanceVariability} and $12.8\%$ for CPU~\cite{SPECCPU2017}).
We still find that some components have significant variance such as memory, CPU cache, and OS-related operations ($4.92\%$, $8.82\%$, $14.39\%$ coefficient of variation, respectively).
Given that significant variability still exists in the cloud, this motivates the question:
\emph{How does platform performance variability affect ML-based tuning of software systems that use these components?}

To explore this, we run a case study tuning \postgres~\cite{postgres} on a well-known workload, \texttt{TPC-C}~\cite{tpcc}, in the cloud.
We use \texttt{SMAC}~\cite{SMAC}, a popular Bayesian Optimizer.
We chose \postgres as our SuT, as, in addition, to disk and CPU, \postgres also heavily uses cache, memory, and the OS - the components which we found have high variance in the cloud.
Surprisingly, we find that \emph{$39.0\%$ of the configs seen during tuning are ``unstable''}; i.e., TPC-C throughput varies substantially, with up to $101.3\%$ CoV.
Worse, many of the configs that performed best during tuning, when deployed to a new VM, experienced up to \emph{$76.1\%$} performance degradation compared with that during tuning.
However, not all best-seen configs experienced such degradation, and there exist configs that are stable and yield high performance.

Motivated by the above findings, we develop \SystemName, a new sampling methodology that aims to find stable, but still high-performing, configs.
\SystemName primarily changes how we measure SuT performance in the \autotuning design, allowing \SystemName to directly integrate with existing optimizers (e.g. \texttt{SMAC}~\cite{SMAC}, \texttt{botorch}~\cite{botorch}, etc.) and generalize to any SuT. 
We use two main insights in \SystemName: (a) the only way to quickly detect unstable configs is by sampling across a representative cluster to capture the variance across nodes, and (b) component-level metrics can mitigate noise from samples.

\SystemName integrates these insights through three main components.
First, we use \emph{multi-fidelity sampling} to evaluate configs at various "budgets", with higher budgets corresponding to sampling from more nodes.
Secondly, we build an outlier detector and an aggregation policy using these additional collected samples that, together, filter out unstable configs.
Finally, we design a \emph{Noise Adjuster} model that aims to remove the influence that platform variability has on the optimizer using component-level metrics (e.g., CPU, Memory stats etc.).
Together these techniques help \autotuning frameworks achieve fast cost-effective convergence to a stable config that is both performant and robust.

We evaluate \SystemName under various scenarios to show that it can generalize well to different scenarios.
These include three SuT (\postgres, \redis, and \nginx), six workloads (\texttt{TPC-C}, \texttt{epinions}, \texttt{TPC-H}, \texttt{YCSB-C}, \changed{\texttt{mssales}~\cite{mssales}, an internal Microsoft production workload, and a Wikipedia serving workload)}, two cloud regions, two distinct VM SKUs, and two optimizers (Bayesian Optimization, Gaussian Processes).
We compare \SystemName against the existing state-of-the-art approach to tuning with a single machine.
Every system and workload evaluated improved when using \SystemName, either in terms of better-achieved performance, reduced variability, or both.
For example, \texttt{mssales}~\cite{mssales}, tuned using \SystemName has $46.8\%$ lower latency and $61.2\%$ lower standard deviation than a system tuned with traditional sampling techniques.

\noindent\textbf{Open Source.}
\added{To benefit the community we release the cloud measurement \emph{dataset}~\footnote{\url{https://aka.ms/mlos/tuna-eurosys-dataset}} and \emph{source code}~\footnote{\url{https://aka.ms/mlos/tuna-eurosys-artifacts}} for \SystemName.}

\section{Background}
\label{sec:background}
\noindent\textbf{Performance Variability.}
Systems performance variability comes in a variety of forms (e.g., inherent software non-determinism, subtle hardware differences, interference from other "noisy neighbor" workloads on shared infrastructure such as in cloud environments, etc.) and has been studied for many decades in many applications of system design under different names (variability, interference, noise, cloud weather, etc.) \cite{TamingPerformanceVariability, CpuCorePerfVariance, HPCVariability, SSDPerfTransperency, GPUVariance}.
Recent work has studied hardware variability that can lead to gray failures \cite{GrayFailure1, GrayFailure2, OriginalGray}.
Many types of systems attempt to be agnostic to absolute system performance or have solutions that do not need config tuning and evaluation.
For example, systems like MapReduce schedule duplicate work on additional machines to avoid stragglers~\cite{MapReduce}.
Other systems~\cite{PARTIES, FIRM, ResourceCentral} attempt to more directly address interference caused by sharing infrastructure.
Similar concepts have also been deployed in the hypervisor~\cite{ppXen} and in the HPC community~\cite{HPCInterference, NoiseInTheClouds}.

In the context of software testing, there has been work to detect outliers caused by performance regression in the cloud rather than the system design.
Some works are as simple as taking more samples across machines~\cite{howbadistesting}, or better test orderings~\cite{testordering}.
Other works have integrated metrics into outlier detection~\cite{outlierdetection1, outlierdetection2}, however, there is no work we are aware of that goes beyond using metrics for detection.

We argue that while further systems advancements in isolation, detection, and scheduling are useful and important, interference will continue to be an issue for systems developers and operators, particularly those working with shared infrastructure, and hence any system that intends to run in the cloud needs to be capable of addressing increased noise.

\noindent\textbf{ML-based Tuning.}
There have been many works that perform ML-based tuning of data systems~\cite{Llamatune, Ottertune, CDBTune, QTune, DBBert, gpttuner}.
These approaches can be either online or offline, and both are sensitive to noise in the environment.

Online approaches use an agent-based approach to change a smaller set of parameters that can be tuned without interrupting the system (e.g., restart).
The policies employed are typically more conservative in order to avoid incorrect configs which may impact the liveness of the system and as such require more tooling and are more challenging to explain and support~\cite{OnlineTune}.
Moreover, since workloads are not replayable, the learning process is more complicated.

Offline tuning, in contrast, allows for a greater degree of exploration.
Given a workload, the configuration space is searched out-of-band on a set of test machines and later periodically applied to the production systems as workloads change.
Since these events can be scheduled and monitored, much like a normal user config change, they are easier to deploy and reason about, and are generally the preferred initial approach by both support teams and customers.
For this reason, we primarily focus on offline tuning in this paper, but many of our lessons apply to both contexts.

In either case, the two primary performance considerations for an autotuner are (a) its ability to \emph{discover} an optimal config, and (b) its rate of \emph{convergence}.
Since the global optimal is typically not known unless one performs an exhaustive (and expensive) grid search, ML-based techniques are used to improve the search rate towards the "best found" config.

Many of the most successful systems use some variant of Bayesian Optimization (BO)~\cite{Ottertune, Llamatune, OtterTuneRealWorld} for their optimizer, as it can often find the optimum in fewer evaluations.
Reinforcement learning (RL)~\cite{QTune, CDBTune} has also been used for this task, yet RL often requires an initial training phase which can be time-consuming.
CDBTune focuses on parallelizing training in a cloud environment, however, it does not address the noise that is inherent in the cloud~\cite{CDBTune}.
We argue that for these systems to be reliable in production, the tuning system needs to account for the variance across machines, changes in collocated VMs, and the noise of the cloud environment.

Prior studies have proposed ML algorithms that are resilient to noisy data~\cite{MLNoiseSensitivity, BOWithNoise, ConstrainedBO}.
There are theoretical studies that characterize, and devise ways to handle constant Gaussian noise in BO~\cite{BOWithNoise, ConstrainedBO}, which make strong assumptions about the shape of the noise and do not apply to the non-parametric noise that we observe in the cloud.
Additionally, such methodology has never been applied to tuning systems.

\vspace{-1ex}
\section{Motivation}
\label{sec:motivation}
Existing approaches for tuning systems have assumed that tuning environments represent eventual deployment environments.
However, it has been reported that even on bare-metal (i.e., isolated) nodes, performance variability can be as high as $29.2\%$ CoV for network, and $16.0\%$ CoV for memory ~\cite{TamingPerformanceVariability, GPUVariance, CpuCorePerfVariance}.
The situation can be even worse in cloud environments, where performance variability due to shared infrastructure has been reported to be even higher~\cite{CloudIOVariance, CloudPerfVariability, CloudPerfVariability2}.
In this section, we investigate how performance variability affects ML-based tuners regarding convergence rates and final configs.
We use a series of experiments understand:
(i) the impact of performance variability (noise) on tuner convergence,
(ii) the magnitude of noise in the cloud, and
(iii) the impacts of cloud noise on best-performing configs.

\subsection{Impact of Interference on Tuner Convergence} \label{sec:cloud_convergence}
\begin{figure}[t]
    \centering
    \includegraphics[width=0.85\linewidth]{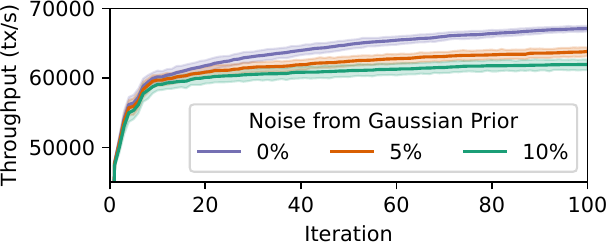}
    \vspace{-2ex}
    \Description[Optimizer Rate of Convergence for epinions workload, running on \postgres 16.1 at various levels of noise.]{A graph showing the difference in rate of convergence depending on the noise taken during a sample.}
    \caption{Optimizer Rate of Convergence for epinions workload, running on \postgres 16.1 at various levels of noise.}
    \label{fig:cloudconvergence}
\end{figure}

We start by investigating the impact of noise on tuner convergence.
We employ \postgres v16.1 as an example SuT we wish to optimize, and we run the \texttt{epinions}~\cite{epinions} workload.
More details about the workload are given in \Section\ref{sec:eval_workloads}.
We use \texttt{SMAC}~\cite{SMAC3} as our optimizer, as done in prior state-of-the-art work~\cite{Llamatune}, and maximize throughput as our objective.
Here, we employ isolated \texttt{c220g5} bare-metal nodes from CloudLab~\cite{Cloudlab}, which do not experience any noise from virtualization or noisy neighbors.
To understand the impact of noise on convergence, we synthetically simulate various levels of noise by reporting noise adjusted using a Gaussian distribution prior.
In particular, if $P$ represents the measured performance, $\sigma$ represents our chosen CoV, then $P^* = P \times \Delta$ is the value we report back to the tuner, where $\Delta \sim \mathcal{N}(1, \sigma^2)$.
We explore three such noise levels (i.e., sampling noise): $0\%$, $5\%$, and $10\%$
We execute $100$ tuning runs, each with a different initialization set, for each level of noise; for each run, we evaluate $100$ samples (i.e., configs).

Figure~\ref{fig:cloudconvergence} depicts the performance of the best config found so far by the tuner at a given iteration.
The solid line is the mean performance, and the shaded region is the $99\%$ confidence interval.
We observe that the tuner convergence rate rapidly degrades when the noise in the prior is increased from $0\%$ to $5\%$.
The tuner using a noise level of $0\%$ converges to the same noise-free performance in $40$ iterations as the $5\%$ noise case in $100$ iterations.
The slow-down ratio of these numbers is called \emph{time-to-optimal} performance~\cite{Llamatune}, and we find it to be $2.50$x.
Increasing noise to $10\%$ further hinders convergence, and gives a \emph{time-to-optimal} ratio of $4.35$x as compared to no noise.
Thus, given that a small amount of \emph{synthetic} sampling noise can have such a significant impact on the tuner convergence, and hence its cost, we must investigate the magnitude of noise in the cloud to determine how much impact it will have on tuning systems.

\TakeawayBox{Increased sampling noise during tuning significantly slows down convergence, increasing tuning cost.}

\label{sec:cloud_noise}
\subsection{Quantifying Cloud Noise}
\begin{table*}[t]
\begin{center}
\resizebox{\textwidth}{!}{
\begin{tabular}{lclcclccccc}
\hline
Paper                                   & Year & Duration  & Samples   & Instances & Platform                   & Disk          & Memory    & CPU       & Network & OS   \\ \hline
\citet{EC2PerfVariance}                 & 2010 & 4 weeks   & 6 k       & 4         & A                        & \cmark        & \cmark    & \cmark    & \cmark  & \xmark    \\
\citet{LongCloudPerfProd}               & 2011 & 52 weeks  & 250 k     & N/A$^a$   & A, G                   & \xmark        & \xmark    & \cmark    & \xmark  & \xmark    \\
\citet{Moreforyourmoney}                & 2012 & 2 weeks   & 59k       & 40        & A                        & \cmark        & \cmark    & \cmark    & \cmark  & \xmark    \\
\citet{IaaSVariance}                    & 2016 & 4 weeks   & 54k       & 82        & A, M, G, I       & \xmark        & \cmark    & \cmark    & \xmark  & \xmark    \\
\citet{TamingPerformanceVariability}    & 2018 & 46 weeks  & 900 k     & 835       & CL                   & \cmark        & \cmark    & \xmark    & \cmark  & \xmark    \\
\citet{PerformanceEvalCloudFunctions}   & 2018 & 22 weeks  & 730 k     & 13,723$^a$ & A, M, G, I       & \xmark        & \xmark    & \cmark    & \xmark  & \xmark    \\
\citet{CloudBenchmarkSuite}             & 2018 & 4 weeks   & 63 k      & 244       & A                        & \cmark        & \cmark    & \cmark    & \cmark  & \xmark    \\ 
\citet{bigdatareproducable}             & 2020 & 3 weeks   & 1000 k    & 1         & A, G, H         & \xmark        & \xmark    & \xmark    & \cmark   & \xmark   \\
\citet{NoiseInTheClouds}                & 2022 & N/A       & 516 k     & 2         & A, M, G, O    & \xmark        & \xmark    & \xmark    & \cmark  & \cmark    \\ \hline
This Work                               & \textbf{2024} & \textbf{68 weeks}  & \changed{\textbf{7037 k}}    & \changed{\textbf{43,641}}     & M                      & \cmark        & \cmark    & \cmark    & \xmark & \cmark    \\ \hline
\end{tabular}}
\\
\raggedright
\footnotesize{$^a$ These studies were conducted on serverless nodes}
\end{center}
\vspace{1ex}
\caption{Comparison of our work to prior benchmarking studies. For platform A indicates Amazon AWS, M indicates Microsoft Azure, G indicates Google GCP, I indicates IBM Cloud, O indicates Oracle OCI, CL indicates CloudLab, H indicates HPCCloud.}
\label{table:CloudStudy}
\vspace{-1em}
\end{table*}

Performance variability in the cloud has been studied in multiple prior works (shown in Table~\ref{table:CloudStudy}); however, we find that these prior studies are limited in scale and duration, or outdated, and do not report results on OS-related variability.
We also note that while some studies (e.g., \citet{PerformanceEvalCloudFunctions}) do sample many instances, they use serverless nodes, which have less consistent performance properties than VMs~\cite{MicroserviceVarianceFromServerless}, as they have weaker isolation.

We thus conduct a study to investigate the magnitude of variability in today's cloud caused explicitly by platform and hardware variability.
We ran a longitudinal study for $480$ days on Microsoft Azure, starting May 28th, 2023, and ending on September 19th, 2024.
We conducted our study using $40$ microbenchmarks, and $13$ end-to-end application benchmarks, resulting in a total of $92$ metrics.
We investigate \texttt{Standard\_D8\_v5}~\cite{D_Series} VMs general-purpose VM in two regions: \texttt{westus2}, and \texttt{eastus} and across two types of lifespans: short-running, and long-running VMs.
Additionally, we briefly discuss \texttt{Standard\_B8ms}~\cite{B_Series} VMs in the same regions and with the same lifespans.
We isolated our benchmarks to use \texttt{SSDv2}~\cite{ssdv2} disks in Azure where applicable.
Long-running VMs ran for the entire duration of the study (68 weeks).
Short-running VMs were deployed, ran the benchmarks, recorded the results, and were immediately deprovisioned.
This allows us to sample across many physical nodes in each region, collecting different types of cloud weather as compared to long-running VMs, which we find seldom migrate.
There were three long-running VMs of each type in each region, and a total of \emph{over $43$k} short-lived VMs approximately evenly distributed between the two regions and types of VM.
During the study, we collected \emph{over $7$M} data points approximately evenly distributed between the two regions, the two VM lifespans, and the two VM types.

\begin{figure}[t]
    \centering
    \includegraphics[width=0.85\linewidth]{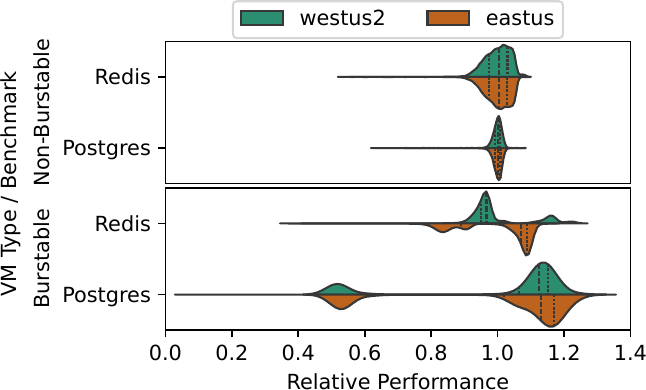}
    \vspace{-1ex}
    \caption{\added{The variance of \postgres and \redis benchmarks. Relative performance is relative to the mean performance seen within a region and VM SKU type. Higher is better.}}
    \label{fig:vm_type_differences}
    \vspace{-2ex}
\end{figure}

First, we compare the end-to-end performance of long-lived burstable VMs to non-burstable VMs running two full systems, \postgres and \redis, in Figure~\ref{fig:vm_type_differences}.
For PostgreSQL, we run a read-write workload from \texttt{pgbench}~\cite{pgbench} with a dataset significantly larger than memory.
For \redis, we run a write-heavy workload from \texttt{redis-benchmark}~\cite{redisbenchmark}.

Immediately, it is clear that there is not only significantly higher variability in burstable VMs but also a bimodal distribution.
There are two key differences between these two types of VMs.
First, burstable VMs are oversubscribed with weaker isolation as compared to non-burstable VMs.
This causes a wider distribution of performance.
Secondly, and more problematic for tuning, bursting credit depletion causes extreme performance bimodality.
For example, when there are sufficient disk credits for the \postgres workload, we see performance in the upper end of the distribution, and when they are depleted, there is a greater than $50\%$ degradation.
This is not only problematic because of the loss of performance, but it also makes it difficult for the optimizer to distinguish between credit depletion and unstable configurations (\S~\ref{sec:unstable_configs}).
For this reason, for the remainder of the paper, we choose to use non-bursting VMs.
We leave accounting for burstable VM credits during tuning as future work.

\TakeawayBox{Burstable VMs are not suitable for autotuning applications without accounting for bursting credit depletion.}

\begin{figure}[t]
    \centering
    \includegraphics[width=0.85\linewidth]{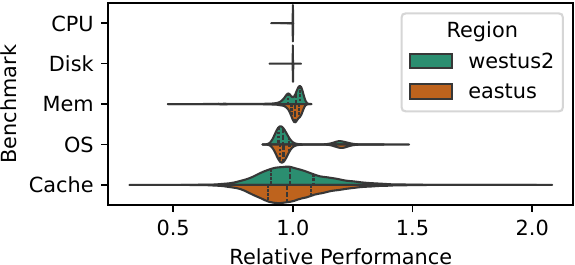}
    \vspace{-1.5ex}
    \Description[Properties of old components.]{A comparison of variance for memory, OS benchmarks, and Cache, i.e., the old components.}
    \caption{The variance of benchmarks targeting CPU, Disk, Memory, the OS, and CPU cache. Relative performance is relative to the mean performance seen. Higher is better.}
    \label{fig:hardware_variability}
\end{figure}

We focus on the results of $5$ different resource microbenchmarks running on non-burstable VMs: prime verification using \texttt{sysbench}~\cite{sysbench} to test CPU performance, random writes using \texttt{fio}~\cite{fio} with the Linux Async IO engine for disk, Max Bandwidth with a 1:1 read/write ratio using Intel's Memory Latency Checker~\cite{IntelMemoryLatencyChecker} for memory, thread creation time using \texttt{OSBench}~\cite{osbench} for the OS, and the \texttt{stress-ng}~\cite{Stressng} cache micro-benchmark.
While these are not the only benchmarks we ran for each of these components, they are each representative of other benchmarks that targeted their respective components.
The results of these microbenchmarks are shown in Figure~\ref{fig:hardware_variability}, and we present two main takeaways.

First, the performance characteristics of some components (e.g., disks and CPUs) have significantly lower variability than the results reported in prior works  (Table~\ref{table:CloudStudy}), with the highest CoV for any disk-isolated benchmark or any CPU-isolated benchmark being $0.36\%$, and $0.17\%$, respectively.
This decreased variability can likely be attributed to changes in the offerings provided by the platform.
Prior studies have shown many different CPUs all within a single VM SKU, including a mix of Intel and AMD CPUs~\cite{Moreforyourmoney}.
Modern VM SKUs, on the other hand, have much tighter SLAs and have separated these CPUs into different SKUs~\cite{D_Series, EC2Offerings, GCPVendorBenchmarks, OracleCloudOfferings}.
In all of our experiments, we found that every VM reported using the same underlying CPU type.
Additionally, there has also been work at the platform level to fairly share and isolate CPU cycles~\cite{CPUIsolationGoogle}.
The virtual disk service also benefits from platform-level improvements.
In Azure, we found that the SSDv2 Managed Disk offering~\cite{ssdv2} has significantly lower variability than storage shown in prior work.
Note that Azure is not the only platform to provide these types of virtual disks.
GCP and AWS both provide similar offerings\cite{AWSDisks, GCPDisks}

\TakeawayBox{Platform level advancements have made CPU and Disk cloud-based components much more stable than results shown in prior studies.}

However, platform-level improvements have not solved all the variability problems that exist in the cloud.
Memory, OS, and Cache still have high variability with CoVs of $4.92\%$, $9.82\%$, and $14.39\%$ respectively.
Some of the variance can be explained by design decisions that cloud providers have made.
Memory and cache bandwidth are often not restricted, which allows heavy interference impacts.
There has been work to more fairly share these resources at a platform level~\cite{MemoryBandwidthParitioning}, and a hardware level~\cite{IntelNFV, IntelDDIO}, however, these have either not been deployed or have not completely resolved the issue.
OS operations are another problem.
Many OS-related operations require a \texttt{VMEXIT}, which can incur a heavy overhead due to CPU overbooking and modern side-channel and other security mitigations that systems must employ \cite{prout2018measuring}.
While recent proposals have introduced some hardware support~\cite{IntelVtX, IntelHDIOV}, we find that they do not fully mitigate variability in virtualized systems.
Given that there is variability in cloud-based systems, it raises the question: \emph{how much does this variability impact tuning?}

\TakeawayBox{Memory and Cache still experience high levels of variance, we also see the OS as a previously unmeasured source of variance.}
\vspace{-1em}

\subsubsection{Unstable Configurations}
\label{sec:unstable_configs}
\begin{figure}[t]
    \setlength{\abovecaptionskip}{1ex}
    \centering
    \begin{subfigure}[t]{0.5\linewidth}
        \includegraphics[width=\linewidth]{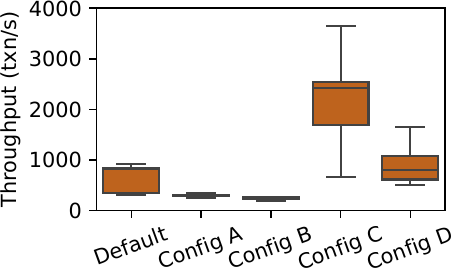}
        \Description[Properties of configurations in the initialization set.]{A selection of the first 10 configurations displaying the difference in variance in the different configurations.}
        \caption{Throughput for the $5$ configurations of the initialization set evaluated on the same $30$ nodes.}
        \vspace{1ex}
        \label{fig:unstable_init}
    \end{subfigure}
    \hspace{0.03\linewidth}
    \begin{subfigure}[t]{0.4\linewidth}
        \includegraphics[width=\linewidth]{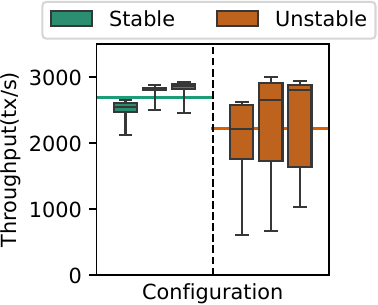}
        \caption{A subset of best-performing configs. when deployed on new nodes.}
        \vspace{1ex}
        \label{fig:best_configs}
    \end{subfigure}
    \caption{The performance of initialization and of best-performing configurations (picked by the optimizer) during training (left) and when deployed on new nodes (right). 
    }
    \label{fig:unstable_tuning}
\end{figure}
To study the impacts of tuning the cloud, we ran $30$ tuning runs, each with the same initialization set, targeting \postgres 16.1 running \texttt{TPC-C}~\cite{tpcc} for $50$ iterations.
We evenly distributed these $30$ runs across $30$ identically configured \Dsv~\cite{D_Series} VMs using SSDv2 Data Disks in the same region.
The question we wish to answer with this experiment is: \textit{are modern tuning processes resilient to the performance variance seen in \Section\ref{sec:cloud_noise}, and in turn always able to generate a usable configuration?}

First, we notice that the configurations in the initialization set have vastly different stability characteristics. Of the $10$ total configurations in the initialization set, we present the $5$ configs that do not immediately crash the system in Figure~\ref{fig:unstable_init}.
Config C is particularly concerning as it either performs extremely well or extremely poorly, depending on which machine it was run on, in a difficult-to-predict manner.
We term these configurations \emph{Unstable}.

Next, we explore the ability of the best-performing configurations to be deployed on a VM distinct from the one used during tuning.
We term the ability to maintain performance as the \emph{transferability} of a config.
To explore \emph{transferability}, we deployed the best config found from each optimization run, to a set of $10$ new VMs with the same platform setup, and evaluated the performance of the optimized config on the new VM.
We continue to find vastly different variability between these "best" configs, with some configs being unstable and others being stable.
We can quite easily divide these configs into the two aforementioned categories and show three of each in Figure~\ref{fig:best_configs}.
We find from this divide that $13$ of the $30$ transferred configs are unstable.
Furthermore, some of these configs, including two of those shown, can have their performance degraded by over $70\%$, and have CoVs of up to $36.3\%$.
This is significantly higher than what can be attributed to simply noisy neighbor effects.
The highest CoV for any benchmark \postgres benchmark in \S~\ref{sec:cloud_noise} was $7.23\%$.
In \Section\ref{sec:design}, we design a heuristic to implement this classification. 

We thoroughly investigated system performance metrics, such as storage throughput and CPU performance via micro-benchmarks.
These revealed (1) no obvious correlations and (2) that the new machines were performing within the small bounds expected to be caused by platform variance and noisy neighbors.
We eventually found the root cause for this performance degradation to be a difference in the query plan selected by the DBMS at run time.

For \texttt{TPC-C}, when the query planner generates candidate plans, the top two candidate query plans for the \join query are predicted to have almost the same execution time.
However, in reality, one plan takes two orders of magnitude longer.
Small changes in the underlying component performance can tip the balance on which plan is selected.
Machines that performed well always selected the high-performing plan, while machines that performed poorly occasionally picked the poor plan due to minor differences in predictions in the cost models.
We found that the knobs that impact if a config may be \emph{Unstable} include \texttt{enable\_bitmapscan}, \texttt{enable\_hashjoin}, \texttt{enable\_indexscan}, \texttt{enable\_nestloop}, but the exact combinations are inconsistent across configs. 
We also found unstable configs in other systems such as \redis and \nginx, but omit their details given space constraints.

A lack of awareness of \emph{Unstable} configs during performance sampling can lead to \emph{Unstable} configs being promoted during the tuning process.
Without mitigation, this can cause the best apparent config to be \emph{Unstable} and can cause significant performance degradation when deployed.

\TakeawayBox{The existence of \emph{Unstable} configs and their impact on tuning, motivates the need for methods to \textit{collect} representative samples, and \textit{detect} anomalous configs}

\section{Design}
\label{sec:design}

\begin{figure}[t]
    \centering
    \includegraphics[width=0.95\linewidth]{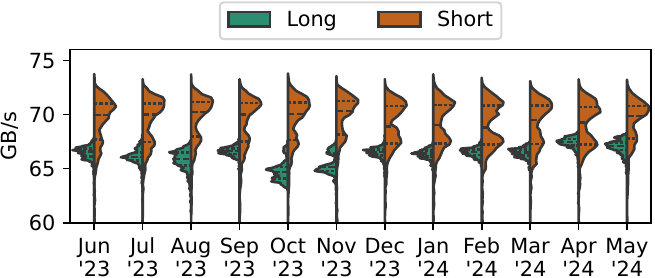}
    \Description[Timeline.]{Timeline.}
    \caption{Performance of a memory bandwidth benchmark partitioned by month. The long-running VM is a single representative VM that ran for the entire duration, while the short-running is a set of $10,632$ VMs, both in \texttt{westus2}.
    }
    \label{fig:timelineperformance}
\end{figure}
\begin{figure*}[t]
    \centering
    \setlength{\abovecaptionskip}{1ex}
    \includegraphics[width=.9\linewidth,page=2]{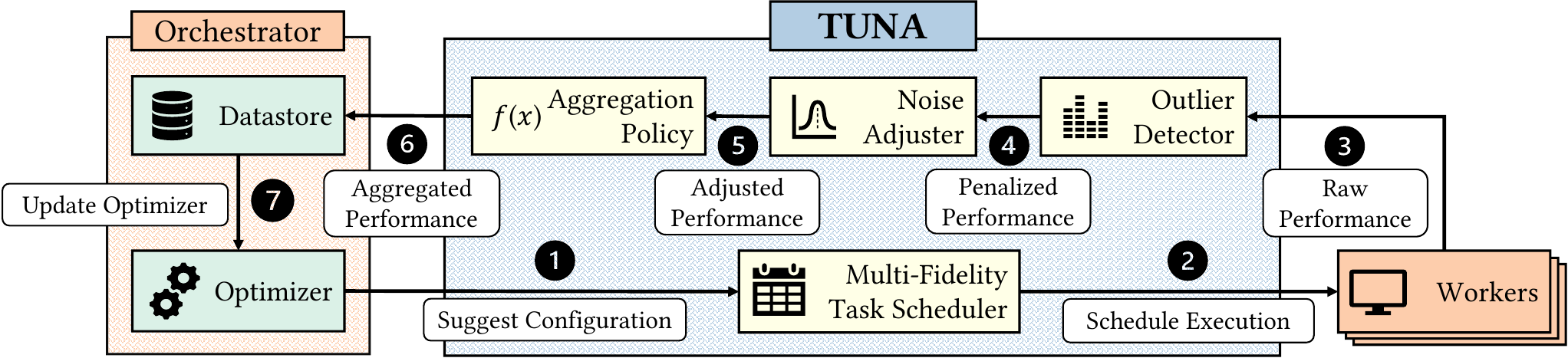}
    \caption{Design for \SystemName with our contributions highlighted lime and blue.}
    \label{fig:systemdesign}
\end{figure*}

The substantial impact of noise on tuner performance and the existence of \emph{unstable} configs (\Section\ref{sec:motivation}) motivates a new system that mitigates the observed noise and produces configs that transfer across different nodes.
Ideally, such a tuning design (i) should be agnostic to the system that is being tuned, (ii) should not require any changes to the underlying optimizer, and (iii) should avoid hurting optimizer performance (e.g., increased time to convergence or degradeyd config quality).

To this end, we present our new tuner design, \SystemName, which achieves the above goals by introducing four key components:
\emph{
(i) Multi-fidelity sampling,
(ii) Unstable configuration detection,
(iii) A model for sample noise mitigation,
(iv) sample aggregation.
}
\SystemName refrains from making any changes to the underlying optimizer or SuT, but rather focuses on changing what data is collected by intelligently distributing samples across machines, augmenting results with system metrics, and unstable configuration detection.
Similar to prior works~\cite{Llamatune, Ottertune, QTune}, \SystemName optimizes a chosen target metric for a static workload.
A high-level overview of how the above components can be fit into an existing tuning setup can be seen in Figure~\ref{fig:systemdesign}.
In the next subsections, we describe these components in more detail.

\subsection{Multi-Fidelity Sampling} \label{sec:multi_fidelity_sampling}

Performance variability across different cloud nodes necessitates evaluating a config on multiple nodes.
Figure~\ref{fig:timelineperformance} depicts the performance of a memory benchmark running on a single long-lived VM for 68 weeks, versus running the same benchmark on many different short-lived VMs over the same period.
We observe that while there are performance changes for our long-lived VM over time, the time scale on which these happen is too long and they do \textit{not} capture the performance variance across the broader cluster on which the config could be deployed.

The naive way for the optimizer to gain more confidence is to evaluate each configuration on many nodes, yet doing so can be prohibitively expensive.
To this end, we leverage \textit{multi-fidelity} sampling.
The key idea of multi-fidelity sampling is to associate a "budget" for each config suggested by the optimizer.
Configs initially begin with a limited budget.
Promising configs are then promoted, reevaluating them with increased budget, increasing the \emph{confidence} that this config is indeed better-performing, without spending "too much" on poor configs.

Multi-fidelity sampling provides a principled way to parallelize config evaluations across multiple nodes and filter bad-performing configs quickly.
We associate the multi-fidelity budget with the number of nodes on which a config is evaluated.
We implement this policy using Successive Halving~\cite{SuccessiveHalving}, a well-known multi-fidelity policy.
This enables us to collect additional samples of individual configs across several machines to evaluate their robustness to noise in the deployment environment.
In particular, we start by evaluating a config on a single VM, followed by a small set of VMs (e.g., $3$).
As long as it keeps performing well, we evaluate it on an increasingly larger set of nodes until eventually, we run it on the entire cluster.
Otherwise, we quickly discard the config to give the evaluation budget to more promising ones, which have not yet been run on as many nodes.

Our multi-fidelity tuning methodology comes with two important benefits concerning optimization robustness.
First, the resulting best-performing configs are more likely to be transferable to a new deployment VM.
This stems from the fact that these configs were already evaluated on multiple nodes, providing us with a more representative performance distribution.
Second, given that we now have a distribution of samples, we can use these to detect \emph{unstable} configs.
We elaborate on how we do this, in the next subsection.

\subsection{Unstable Configuration Detection} \label{sec:outlier_detector}
We recall (from \S\ref{sec:unstable_configs}) that we observed a wide gap in stability between stable and unstable configs.
Given the existence of this wide gap, we introduce a simple heuristic, which, given a set of gathered samples (i.e., performance) for a config, decides whether it is stable or not (i.e., outlier).

Our heuristic should make sure that when configs are classified as stable, they are also \emph{transferable}.
The classification decision is made using the performance observed when evaluating the configuration on multiple nodes.
To this end, the heuristic should only consider whether an outlier performance value exists, i.e., neither the degradation size nor frequency of outliers are important.
For example, a configuration that resulted in a single outlier when run in our tuning cluster does not imply being inherently more reliable than a configuration that had two extreme outlier performance numbers; both should be classified as unstable.

One might try to use, as heuristics, the standard deviation or CoV to classify unstable configs.
Yet, both options are unsuitable.
The former is sensitive to the absolute performance numbers, thus requiring manual tuning for each target system and workload.
The latter is inherently biased towards the ratio of outliers to non-outlier measurements, making it hard to reason about the magnitude of performance degradation.
To this end, we choose \emph{relative range} as our target heuristic for the outlier detector.
This metric does not require tuning and is not biased towards the incidence of outliers.

\[ \text{Relative Range} = \frac{max(x) - min(x)}{\mathbb{E}(x)} \]

Using the above heuristic and a fixed threshold of $30\%$, classify a configuration as stable or not.
If we detect an unstable configuration, we inject a penalty score to prevent the optimizer from searching this region again.
We implemented this penalty as halving the reported performance as in prior work~\cite{OtterTuneRealWorld}, though other policies are also possible.

\begin{figure}[t]
    \centering
    \includegraphics[width=.85\linewidth]{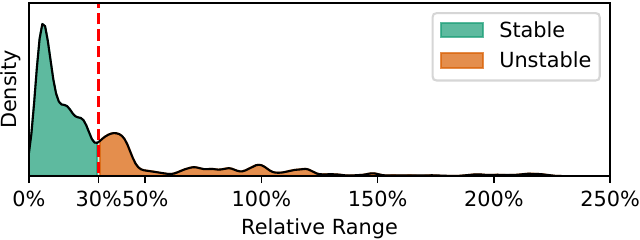}
    \vspace{-1ex}
    \caption{\added{Sensitivity analysis of 1000 configurations seen during tuning. Using a red dashed line, we specify our detection threshold at $30\%$ between the first and second peaks.}
    }
    \label{fig:outlier_sensitivity}
\end{figure}

We justify our detection threshold by analyzing relative ranges from $1000$ configurations seen during tuning, all of which were run on $10$ nodes.
The results of this experiment can be seen in Figure~\ref{fig:outlier_sensitivity}.
We pick a simple threshold value of $30\%$, as this sits in the trough between the first and second peaks of relative ranges.
While there is still significant density in the second peak, we are not concerned about a false positive, as we can always find another well-performing stable configuration.
However, a false negative can be disastrous when deployed, particularly those with extreme ranges, and thus we opt for a slightly conservative value.
This all said the exact choice of this value is not particularly important as long as we safely remove extremely unstable configurations at the far end of the spectrum (right side of the graph), and believe any value between $15\%$ and $30\%$ is reasonable.

\subsection{Sampling Noise Modeling} \label{sec:noise_model}

Aside from detecting unstable configurations, we also wish to mitigate the inherent cloud noise from our measurements to improve the convergence rate (\Section\ref{sec:cloud_convergence}).
Specifically, given the set of noisy application performance measurements, we wish to predict the mean \textit{noise-less} performance value to give the optimizer a more stable signal suitable for learning.
To do this we make use of guest OS resource usage metrics.
Our main observation is that since configs \emph{and} noise change the end-to-end application performance, they also impact the use of system resources in the VM guest OS and can thus be used as an additional signal.
To this end, we propose building a predictive model that given an input noisy performance measurement and a set of system metrics can ultimately predict the noise-less performance.
This is possible as we evaluate a config on multiple nodes via multi-fidelity tuning, giving us additional data that we can leverage for training our predictive model.
Note that while such predictive approaches have been utilized for performing outlier detection~\cite{NoisyNeighborDetection1, NoisyNeighborDetetion2}, they have not been used in the context of mean performance prediction or for systems tuning.

\begin{figure}[t]
\vspace{-3ex}
\begin{algorithm}[H]
\caption{Model Training}\label{alg:modeltrain}
\changed{
\begin{algorithmic}[1]
\Require $C$ \Comment{The configuration evaluated}
\Require $W$ \Comment{A set of workers}
\Require $M_{CW}$ \Comment{Metrics of Configuration run on Worker}
\Require $P_{CW}$ \Comment{Performance of Configuration run on Worker}

\State $X \gets \{M_{cw} \cup \text{OneHot}(w)\} \quad \forall c \in C, \forall w \in W$
\State $y \gets \frac{P_{cw}}{\mathbb{E}[P_{cw} | C=c]} - 1 \quad \forall c \in C, \forall w \in W$ \Comment{Percent Error}

\State $\text{Model} = \text{RandomForestRegressor} \circ \text{Standardize}$
\State $\text{Model}.\text{fit}(X, y)$
\Ensure $\text{Model}$

\end{algorithmic}
}
\end{algorithm}
\vspace{-5ex}
\end{figure}
\begin{figure}[t]
\vspace{-3ex}
\begin{algorithm}[H]
\caption{Model Inference}\label{alg:modelinference}
\changed{
\begin{algorithmic}[1]
\Require $\text{Model}$ \Comment{The model used to predict performance}
\Require $M_{cw}$ \Comment{Metric for a configuration on a worker}
\Require $w$ \Comment{Worker ID}
\Require $P_{cw}$ \Comment{Performance of configuration run on worker}
\Require $O_c$ \Comment{If configuration has outliers}

\State $s \gets \text{Model.predict}({M_{cw} \cup \text{OneHot}(w)})$
\State $m(s, p, o) :=
    \begin{cases} 
      p & o = True \\
      \frac{p}{s + 1} & o = False 
   \end{cases}$
\Comment{Apply Model}

\Ensure $\hat{P}_{cw} \gets m(s, P_{cw}, O_c)$
\end{algorithmic}
}
\end{algorithm}
\vspace{-5ex}
\end{figure}

There are a few important design decisions we make for our noise adjuster model.
First, and most importantly, we do not bring any data from any prior tuning run, \emph{i.e., the model is initialized randomly every time we begin a new tuning run}.
We do this to make comparisons more general to new workloads, similar to how the surrogate model in the optimizer is not updated with prior data.
We leave transfer learning from prior observations to warm up the model for future work.

We use system metrics and a one-hot encoding of the machine ID collected from the configurations with the most samples within a run as training data to build our model.
We do not make any effort to select a subset of metrics (e.g., by using a pre-processing technique), but rather provide all available metrics from \texttt{psutil}~\cite{psutil} to our model.
For our target metric, we use performance relative to the mean performance of a given configuration (Line 4 of Algorithm~\ref{alg:modeltrain}).
This allows our model to learn which metrics are important rather than requiring us to hand-select them per SuT.

Secondly, we require three properties from our model.
(i) The model needs to be able to generalize well over unseen data as we cannot expect configurations to take the same code paths.
(ii) The model must be able to select important input features (i.e., \texttt{psutil} metrics), from a large feature space, without sacrificing predictive performance as we do not wish to encode this information.
(iii) And finally, the model must be able be able to be trained on a small amount of data, as optimization is a cold-start process.
We choose to use a random forest regressor as our predictive model as it has been shown to satisfy all of these properties~\cite{RandomForestBenchmark}.

Finally, we only use data from configs that have been run at the highest budget.
This is the most reliable data for training input to model as it has the highest chance of catching and removing unstable configs, which can negatively impact our model accuracy.
While this limits the amount of training data, we find that the model still converges quickly to produce more accurate predictions.
Additionally, as training random forest regressors is cheap, we simply rebuild the entire model every time there is an added data point.
A description of this entire training process can be seen in Algorithm~\ref{alg:modeltrain}.

We then use this model to predict the true mean performance for every sample that we collect from stable configs.
If we detect an unstable configuration, we bypass the model as these fall outside of our training data, and will already be heavily penalized by the outlier detector.
The inference phase of this model can be described in Algorithm~\ref{alg:modelinference}.
We simply collect the metrics and the machine ID that the configuration was run on and adjust the score we report to the next stage of our sampling process by the performance prediction.

\subsection{Sample Aggregation} \label{sec:aggregation_func}
Finally, we need to generate a single value for the optimizer, and hence we need an aggregation policy.
While it may seem intuitive to use median or mean as an aggregation policy, these both have the potential to ignore outliers.
Additionally, we find that during tuning, when unstable configurations performed poorly (e.g. Config C in Figure~\ref{fig:unstable_init}), the final configurations were not unstable as seen in Figure~\ref{fig:best_configs}.
For this reason, we select \texttt{min} as our aggregation policy, as it correctly penalizes unstable configurations, and also optimizes for the worst case.
While it is still possible for this aggregation policy to have unpredictable performance above the worst case, the outlier detector (\S~\ref{sec:outlier_detector}) bounds this uncertainty to $30\%$.
Future work may explore more complicated parametric aggregation policies that balance optimizing for reduced variance \emph{and} increased performance simultaneously.

\section{Implementation}
\label{sec:implementation}
We implement all the TUNA components in Python.
By default, we use \texttt{SMAC}~\cite{SMAC3} as our optimizer as it supports multi-fidelity sampling natively and has been shown to perform better than other optimizers in the context of system configuration tuning~\cite{FacilitatingDBMSTuning}
We show, however, that \SystemName also supports other optimizers in \S\ref{sec:ablation}.
To manage our SuT we use Nautilus~\cite{Nautilus}, a system that manages instantiation and cleanup for benchmarking a given workload~\cite{krothmlos}.
For our evaluation, we run our sampling method for the same amount of time as a traditional single-machine methodology.
The rest of this section focuses on some specific implementation decisions used in our evaluation.

\subsection{Cluster Size}
\begin{figure}[t]
    \centering
    \includegraphics[width=.85\linewidth]{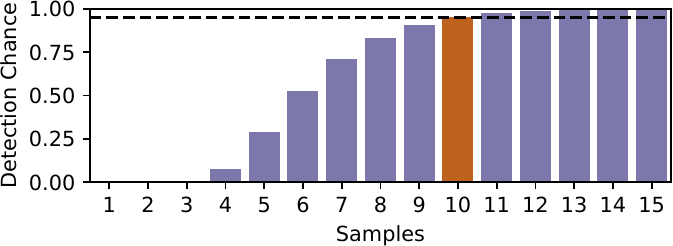}
    \vspace{-1ex}
    \Description[Chance of detecting all unstable configurations.]{The chance of detecting all unstable configurations in a tuning run based on the number of nodes it was run on. This result is based on the data collected in Section~\ref{sec:unstable_configs}.}
    \caption{The chance of detecting all unstable configurations in a tuning run based on the number of nodes it was run on. This result is based on the data collected in Section~\ref{sec:unstable_configs}.}
    \label{fig:detection_chance}
\end{figure}

One important decision is choosing how to manage the nodes that are required for Multi-Fidelity Sampling (\Section\ref{sec:multi_fidelity_sampling}).
One option would be to provision a new node for every sample, however this has high overheads from spinup and spindown.
Instead, we use a fixed cluster.
Determining the size of the cluster is a trade-off between confidence in detecting an unstable configuration as unstable and sample efficiency.
While increasing the number of nodes in the system will increase the parallelism, at the same time it will lead to higher costs and increased delays in obtaining samples with the maximum budget. 
To combat this, we choose to make the cluster as small as possible to give us $95\%$ confidence that we will detect all unstable configurations based on the unstable configurations described in \S~\ref{sec:unstable_configs}.
We used this data to calculate the average chance of detecting a known unstable configuration when sampling from a given number of nodes.
We then use this to find the odds of all unstable configurations being found with a given cluster size during an entire tuning run.
The results are shown in Figure~\ref{fig:detection_chance} and show that a cluster size of $10$ is large enough to detect unstable configurations with $95\%$ confidence.
Thus, we set the maximum budget as $10$ in our implementation.

Given a fixed cluster size, we need to decide how to place sampling tasks across nodes.
In our implementation, we reuse old samples taken at a lower budget when we need to sample at a higher budget. 
This means that we can decrease the cost of sampling at a higher budget, however, we must ensure that the new samples are not performed on the same nodes as the prior runs to ensure the detection guarantees.
This is done by maintaining a queue of pending configurations that need to be run, where configurations are placed when an eligible node becomes available.
For example, if the optimizer suggests a config at a budget of 3 that was already run at a budget of one on node 1,  the two additional required runs will wait until a different node is free.

\subsection{Example Pipeline} \label{sec:pipeline}
\begin{figure}
    \centering
    \includegraphics[width=\linewidth]{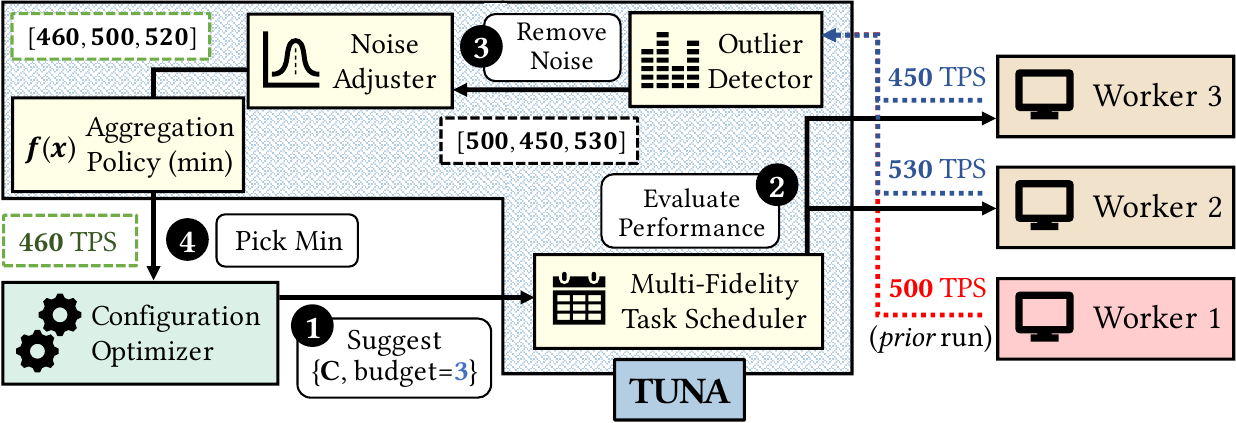}
    \caption{Example Pipeline for TUNA}
    \label{fig:tuna-example}
\end{figure}

Figure~\ref{fig:tuna-example} shows a concrete example of an iteration through the pipeline as follows:
The optimizer suggests a config with a budget of 3, which we have previously run with a budget of 1.
We find that the config was previously run on worker 1, with a throughput of 500 TPS.
Next, we schedule this config to run on \emph{two} workers other than worker 1.
We then receive the results, of 450 TPS, and 530 TPS.
We can then feed all three values (including the prior value) to the outlier detector, which will find that it is a stable config as the relative range ($16.2\%$) is below $30\%$.
Next, as the config is stable, the noise adjuster model adjusts these values to 460, 500, and 520 TPS.
Note that if this is the first iteration, we skip the noise adjuster model, as we have not yet seen any training data.
Finally, we take the minimum and report 460 TPS back to the optimizer as a result.

\section{Evaluation}
\label{sec:evaluation}
\begin{figure*}[ht]
    \setlength{\abovecaptionskip}{1ex}
    \centering
    \begin{subfigure}[t]{.47\linewidth}
      \includegraphics[width=\linewidth]{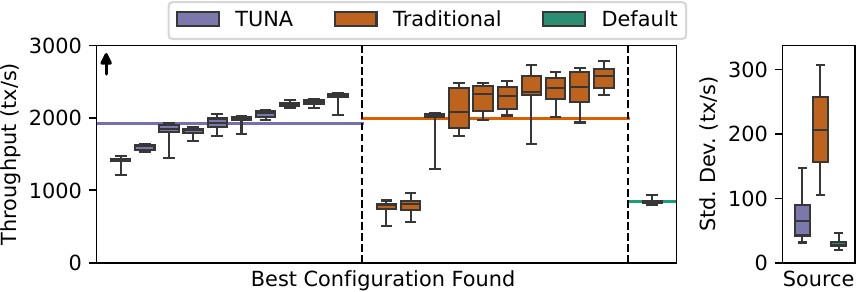}
      \caption{\postgres running \texttt{TPC-C}. Higher Throughput is better.}
      \vspace{1ex}
      \label{fig:eval_tpcc}
    \end{subfigure}
    \hspace{.04\linewidth}
    \begin{subfigure}[t]{.465\linewidth}
      \includegraphics[width=\linewidth]{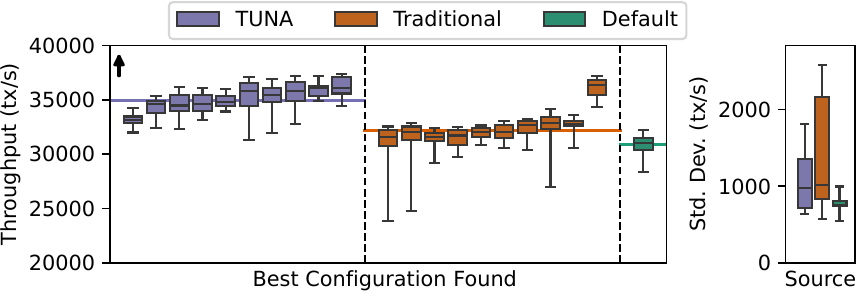}
      \caption{\postgres running \texttt{epinions}. Higher throughput is better.}
      \vspace{1ex}
      \label{fig:eval_epinions}
    \end{subfigure}
    \begin{subfigure}[t]{.47\linewidth}
      \includegraphics[width=\linewidth]{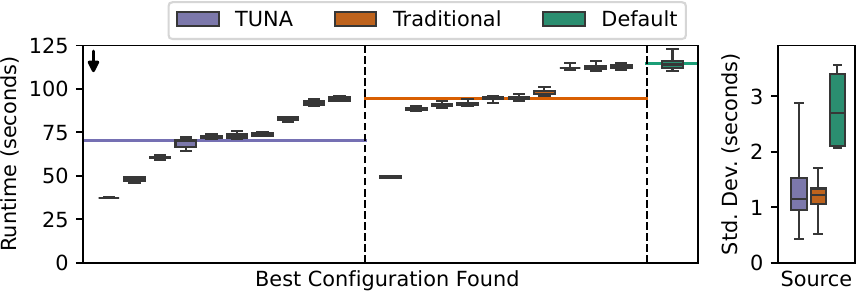}
      \caption{\postgres running \texttt{TPC-H}. Lower Runtime is better.}
      \vspace{1ex}
      \label{fig:eval_tpch}
    \end{subfigure}
    \hspace{.02\linewidth}
    \begin{subfigure}[t]{.48\linewidth}
      \includegraphics[width=\linewidth]{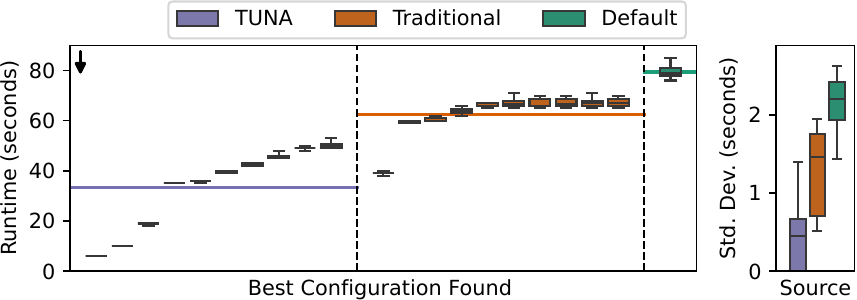}
      \caption{\postgres running \texttt{mssales}. Lower Runtime is better.}
      \vspace{1ex}
      \label{fig:eval_mssales}
    \end{subfigure}
    \caption{Results of applying tuned \postgres configs with several workloads to new systems. The horizontal lines are mean throughput (or runtime).
    In all cases, \SystemName either improves performance, reduces variability, or both.
    }
\label{fig:eval_workloads}
\end{figure*}

In this section, we evaluate \SystemName's ability to effectively find high-performance configurations, as well as reduce variance and prevent unstable configurations.
This means we focus on testing in a way that a real production system would be used: running a tuning methodology offline and then deploying it onto a set of new systems to measure the distribution of performance that a production system could experience.
Specifically, we run the best configuration found during tuning on $10$ new systems and report the results.

For all our evaluations, during tuning and testing, we use a cluster of 10 worker nodes to measure config performance and 1 orchestrator node, totaling 11 nodes.
Unless stated otherwise, all worker nodes use the \Dsv VM SKU with an SSDv2 data disk attached.
All systems evaluated are mounted exclusively on this disk.
For our optimizer, we use SMAC with a random forest surrogate model, except in \Section\ref{sec:ablation}, where we use a \emph{Gaussian Process} optimizer as in OtterTune~\cite{Ottertune}.
This is the same setup that we used in \Section\ref{sec:unstable_configs}.
The orchestrator node is another \Dsv VM without a data disk attached, as the performance variability of the optimizer does not impact the suggestions created by the optimizer. 

For our experiments, we compare against the sampling technique used in prior state-of-the-art works~\cite{Ottertune, Llamatune, CDBTune, QTune}: a single-node sequentially evaluating suggested configurations, without any repeated samples.
We will refer to this as \emph{traditional sampling}.
Both traditional sampling and \SystemName are evaluated for $8$ hours. 
Additionally, we compare \SystemName and traditional sampling with the default (untuned) configuration.
To compare these three, we compare the performance and standard deviation of performance during deployment, not tuning.
Transactional Workloads (OLTP) and workloads minimizing latency are evaluated for $5$ minutes (as in prior work). Analytical workloads (OLAP) target minimizing runtime and have a workload-dependent runtime. 

\subsection{Generalizability Across Workloads} \label{sec:eval_workloads}
First, we fix the SuT, and vary the workload to show that \SystemName can generalize across workloads.
For our first system, we choose \postgres 16.1 as DBMS \autotuning has been a hot area of work, and has already been demonstrated to be a good candidate for \autotuning.
For workloads, we select TPC-C~\cite{tpcc}, and TPC-H~\cite{tpch} as in prior work~\cite{Ottertune}, and additionally evaluate epinions~\cite{epinions}, and \texttt{mssales}, a production OLAP workload from Microsoft.

\noindent\textbf{OLTP workloads.} 
First, for evaluating OLTP workloads, we start with \texttt{TPC-C}, as seen in Figure~\ref{fig:eval_tpcc}.
There is one \join query, and otherwise the rest of the transactions consist of simple single table queries.
For \texttt{TPC-C}, \SystemName achieves $1925$ TPS on average, a $127\%$ improvement over the default, while the traditional sampling setup achieves $1989$ TPS on average, a $134\%$ improvement over the default.
While traditional sampling finds slightly higher performance on average, looking towards the low-performing traditional sampling configurations, two runs perform worse on average than the default.
These two configurations were selected because they performed well (1744 TPS and 2040 TPS) during training.
This is one of the major risks of traditional sampling on single nodes.
Additionally, we see that traditional sampling is significantly more likely to find unstable configurations.
Configurations found using \SystemName have an average standard deviation of $69.0$ TPS, whereas configurations found using traditional sampling methods have an average standard deviation of $205.7$ TPS, a $198.1\%$ increase.

Next, we evaluate \texttt{epinions}, a transactional workload that represents a blog website.
The queries in \texttt{epinions} are simpler than those in \texttt{TPC-C}, but \texttt{Epinions} experiences many of the same problems as TPC-C.
In general, we find that \SystemName achieves $34957$ TPS on average, a $13.2\%$ improvement over the default, while the traditional sampling setup achieves $32189$ TPS on average, a $4.2\%$ improvement over the default.
Convergence to the optimal configuration is much slower than for TPC-C when there is variability, as seen in \S~\ref{sec:cloud_convergence}.
This workload allows us to show the benefits of the noise adjuster model in converging to a higher performance level than in traditional sampling.

Additionally, we find that $3$ configurations learned from traditional sampling are unstable, with a standard deviation greater than $2000$.
Once again, these configurations all performed above $37,000$ TPS during tuning, but the instability was missed by only sampling from one node.
While there is one configuration from TUNA that had higher variability, the CoV was only $5.15\%$, and in the worst case, the performance seen was still $1.46\%$ better than the mean performance of the default configuration.

\noindent\textbf{OLAP workloads.} 
We next move on to OLAP workloads where we try to minimize runtime.
We start by examining TPC-H in Figure~\ref{fig:eval_tpch}, an analytical workload with many, relatively easy, \joins~\cite{job_benchmark}.
Here, we optimize for workload completion time, a realistic metric for analytical workloads.
For \SystemName, on average we can complete the workload in $70.3$ seconds, a $38.6\%$ improvement over the default, while the traditional sampling setup achieves $94.5$ TPS on average, a $17.3\%$ improvement over the default.
In relative terms \SystemName, on average, finds configurations that are $38.4\%$ faster on average with $53.4\%$ lower standard deviation compared to the default.
The standard deviation of \SystemName and traditional sampling methodologies were found to be: $1.3$ and $1.2$ respectively -- both extremely low values.
We believe that this in large part is simply because there are no unstable configurations seen that were optimal for this workload, and the variability simply comes from the platform.
Overall, we can show that \SystemName can converge to a faster configuration without significant changes to stability.

Finally, we present \texttt{mssales}, a production analytical workload, with many complex \join, to show that our techniques are not limited to synthetic workloads.
As seen in Figure~\ref{fig:eval_mssales}, \SystemName, on average, finds a configuration that completes the workload in $33.2$ seconds, with a standard deviation of $0.49$ seconds.
Traditional sampling, on the other hand, on average, finds a configuration that completes the workload in $62.5$ seconds, with a standard deviation of $1.26$ seconds.
In relative terms, \SystemName finds a configuration that is $47.8\%$ faster with $61.2\%$ lower standard deviation, however, it is important to note that the standard deviation is again quite small for both systems.
Overall, this result is quite impressive as the default configuration, on average, takes $79.4$ seconds.
This means that the single-node system finds only marginal improvements, where \SystemName can find significant improvements, sometimes as high as \emph{$2.39$x} speedup.

\subsection{Generalizability Across Regions} \label{sec:eval_region}
\begin{figure*}[t]
    \setlength{\abovecaptionskip}{1ex}
    \centering
    \begin{minipage}[t]{.47\linewidth}
      \includegraphics[width=\linewidth]{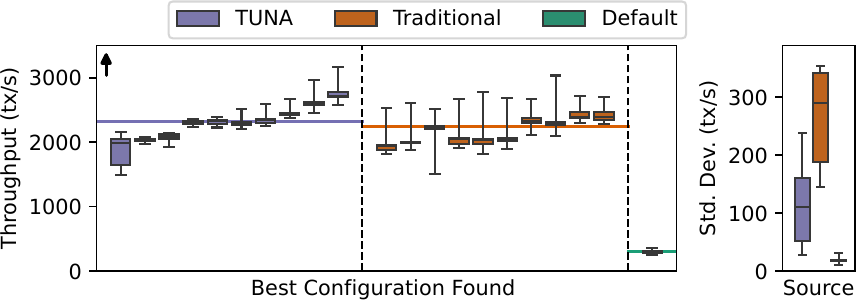}
      \captionof{figure}{
      Results of applying tuned \postgres configs running \texttt{TPC-C} in a new region.
      Higher throughput is better.}
      \label{fig:eval_cus}
    \end{minipage}
    \hspace{.04\linewidth}
    \begin{minipage}[t]{.465\linewidth}
      \includegraphics[width=\linewidth]{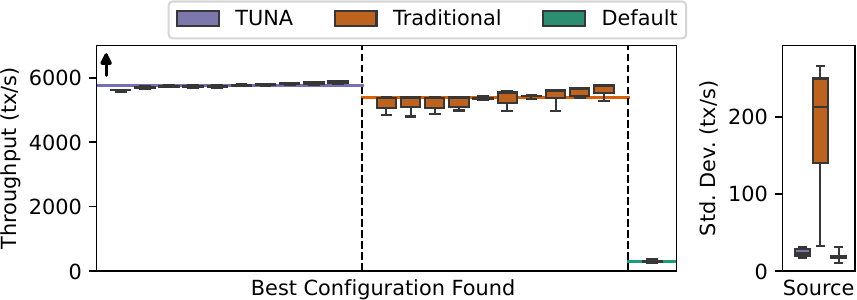}
      \captionof{figure}{Results of tuned \postgres configs running \texttt{TPC-C} on bare-metal CloudLab Nodes.
      Higher is better.}
      \label{fig:eval_cloudlab}
    \end{minipage}
\end{figure*}

Next, to show that \SystemName generalizes across regions with different variability characteristics, we repeat the evaluations in the \texttt{centralus} azure region.
We continue to tune \postgres 16.1, and run \texttt{TPC-C}.
We report the results of this experiment in Figure~\ref{fig:eval_cus}, 
Notably, the variability in this region is higher than the results shown in Figure~\ref{fig:eval_tpcc} in the sense that there are fewer high-performing machines, as shown by the large gaps between the upper quartile and the maximum throughput in the boxplots, which did not exist in the prior experiments.
Empirically, we find \SystemName achieves $2321$ TPS on average, a $669\%$ improvement over the default, while the traditional sampling setup achieves $2239$ TPS on average, a $642\%$ improvement over the default.
Even with the higher variability, \SystemName still mitigates the variability better than traditional sampling, with the average standard deviation of \SystemName being $113.0$ TPS, and traditional sampling being $267.7$ TPS, a $57.8\%$ improvement.

\begin{figure*}[t]
    \setlength{\abovecaptionskip}{1ex}
    \centering
    \begin{minipage}[t]{.47\linewidth}
      \includegraphics[width=\linewidth]{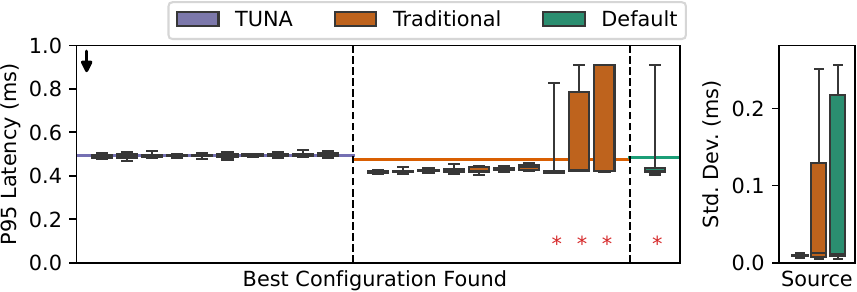}
      \captionof{figure}{\redis running tuned \texttt{YCSB-C} workload configs. 
      Lower is better.
      A red asterisk indicates that a node crashed.}
      \label{fig:eval_redis}
    \end{minipage}
    \hspace{.04\linewidth}
    \begin{minipage}[t]{.465\linewidth}
      \includegraphics[width=\linewidth]{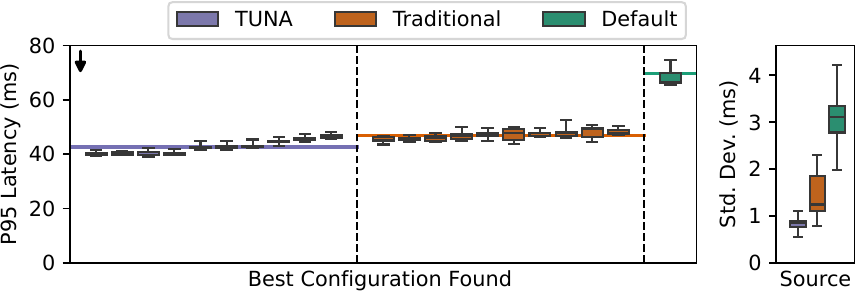}
      \captionof{figure}{Results of applying tuned \nginx configs serving the top 500 Wikipedia pages.
      Lower is better.}
      \label{fig:eval_nginx}
    \end{minipage}
\end{figure*}

\subsection{Generalizability Across Hardware} \label{sec:eval_hardware}
To show \SystemName generalizes across platforms, we repeat the evaluation on CloudLab~\cite{Cloudlab}, a research platform providing bare-metal nodes.
We continue to tune \postgres 16.1, and run \texttt{TPC-C} on \texttt{c220g5} nodes.
As seen in Figure~\ref{fig:eval_cloudlab}, \SystemName achieves $5756$ TPS on average, a $19.1$x improvement over the default, while the traditional sampling setup achieves $5380$ TPS on average, a $17.8$x improvement over the default.
Part of the reason for this extreme performance improvement is that the default configuration does not fully utilize the system memory.
We find that traditional tuning finds many configurations with $7.71x$ higher standard deviation than \SystemName.
Also, we can see that $8$ out of $10$ of these configurations found using traditional sampling are unstable, while all of the configurations found using \SystemName are exceptionally stable and, on average, perform $7.0\%$ better than those found using traditional sampling.

\subsection{Generalizability Across Systems} \label{sec:eval_systems}
To show that our system generalizes across different scenarios, we next perform experiments across two new SuT: \redis, and \nginx.
Neither of these two systems natively supports any of the workloads that we tested in the previous section, so we also introduce new workloads.


For \redis, we use \texttt{YCSB-C}~\cite{YCSB}, a well-known read-only workload with a Zipfian distribution.
Here, we target 95th percentile latency as \redis is often chosen to minimize latency rather than maximize throughput.
With this system, $3$ configurations found using traditional sampling crashed \redis $30\%$ of the time on average.
Furthermore, the default configuration crashed \redis $8\%$ of the time.
These system crashes were caused by an out of memory exception, caused by an overly aggressive configuration.
For these crashed runs, we follow methodology from \cite{OtterTuneRealWorld}, and replace them with the worst P95 latency seen ($.908ms$) on the default configuration as a conservative penalty, rather than $\infty$, mostly for visualization purposes.

With this data-cleaning step, we find that \SystemName has $1.7\%$ higher latency than the default, and $4.3\%$ higher latency than traditional sampling on average. However, in this case the main benefit of \SystemName is that \emph{no configurations crashed}.
This results in \SystemName, on average, having $27.5\%$ lower standard deviation than the default, and $86.8\%$ lower standard deviation than traditional sampling.

For \nginx, we use a custom workload serving Wikipedia.
We serve the Top 500 Wikipedia pages in 2023, including all media content, in the same distribution as these pages were accessed over this same period.
The reported latency is the latency to serve the entire page, including all
We target 95th percentile latency, as web servers often target minimizing high percentile latencies.
As seen in Figure~\ref{fig:eval_nginx}, on average, the 95th percentile latency of the workload with TUNA is $42.6$ ms, a $38.9\%$ improvement over the default, while the traditional sampling setup achieves $46.6$ ms, a $32.7\%$ improvement over the default.
We also find that the standard deviations for \SystemName and traditional sampling are $0.82$ ms and $1.46$ ms, respectively.
\SystemName has a standard deviation that is $63.3\%$ lower than when using traditional sampling.

\subsection{Equal Cost Comparisons}
Previously, throughout the evaluation, we compared TUNA against traditional sampling in the setting where both approaches take equal time.
An alternative to this is tuning with equal cost.
This means that each sampling policy should use an equivalent number of samples.
There are two main ways that an equal-cost budget could be implemented.
First, one could simply extend the traditional sampling technique, as described at the start of \S\ref{sec:evaluation}, to run for more iterations, to match the number of samples used by \SystemName.
Alternatively, one could use a cluster for tuning, and run each configuration on every node in the cluster (equivalent to max budget in \SystemName), and aggregate the result for the optimizer.
To compare \SystemName against both these equal-cost baselines, we again tune \postgres running \texttt{TPC-C} on Azure.

\subsubsection{Extended Traditional Sampling}
\begin{figure}[t]
    \centering
    \includegraphics[width=\linewidth]{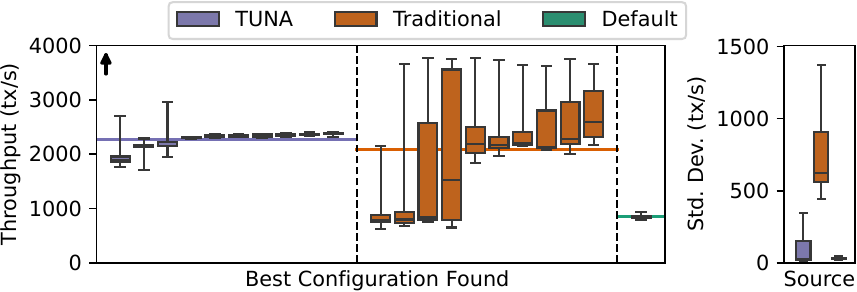}
    \Description[Single node with equal budget.]{Noise Adjuster Model Component Analysis.}
    \caption{\added{Results of applying tuned \postgres configs running \texttt{TPC-C} comparing \SystemName and traditional sampling when budgets (i.e., costs) are both 500. Higher is better.}}
    \label{fig:eval_single_cost}
\end{figure}
Simply extending traditional sampling exacerbates the problem of instability previously seen (Figure~\ref{fig:eval_tpcc}).
Examining the results shown in Figure~\ref{fig:eval_single_cost}, we can see that while the maximum performance is higher, the variability also becomes even higher.
We can see that in this case, because of the extreme performance degradation seen on nodes during deployment using traditional sampling, \SystemName's average performance is now $9.2\%$ higher than in traditional tuning with $87.8\%$ lower standard deviation.

\subsubsection{Naive Distributed Sampling}
\begin{figure}[t]
    \centering
    \includegraphics[width=.425\linewidth]{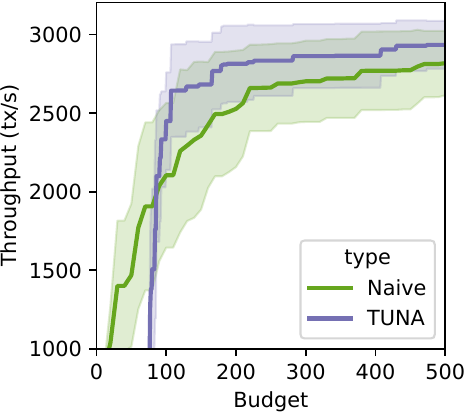}
    \includegraphics[width=.4\linewidth]{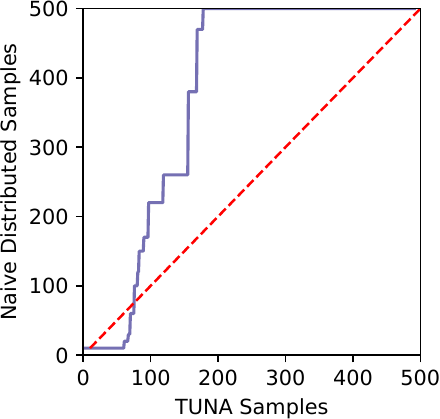}
    \Description[Convergence Comparison of Naive to TUNA.]{Convergence Comparison of Naive to TUNA.}
    \caption{\added{Results of tuning \postgres running \texttt{TPC-C} comparing \SystemName to naive distributed implementation. Plot on the left shows convergence rates in terms of absolute performance. Plot on the right shows convergence gain comparisons where above the dashed line implies \SystemName converges faster, and below the line, naive distributed is better.}}
    \label{fig:eval_naive}
\end{figure}

The alternative of simply running every configuration on every node causes extremely slow convergence.
As an aggregation policy is now required to report a single score to the optimizer, we will use the same aggregation policy as \SystemName, \texttt{min}.
To compare convergence rates, we compare the number of samples \SystemName takes to achieve the same performance as a naive distributed implementation.
Initially, we see that \SystemName converges slower than the Naive Distributed implementation.
This is because both systems only compare configurations run at the maximum budget.
The multi-fidelity sampling technique used in \SystemName does not evaluate at higher budgets until sufficient configurations have been evaluated at a lower budget.
Once \SystemName begins to evaluate configurations at the maximum budget, at around 100 samples, \SystemName quickly jumps ahead of the performance of the naive distributed implementation.
The final result ends with \SystemName achieving the same performance within $206$ samples on average, or in other words, \SystemName converges $2.47$x faster.
If \SystemName is allowed to continue running until $500$ samples, it will achieve $4.1\%$ higher performance.
This is not higher in large part because there are significant diminishing returns to tuning for longer.
Prior work has discussed intelligent stopping criteria, however, this is orthogonal to our approach~\cite{tuneornot}.

\subsection{Component Analysis} \label{sec:ablation}
\noindent\textbf{Optimizer.}
\begin{figure}[t]
    \centering
    \includegraphics[width=\linewidth]{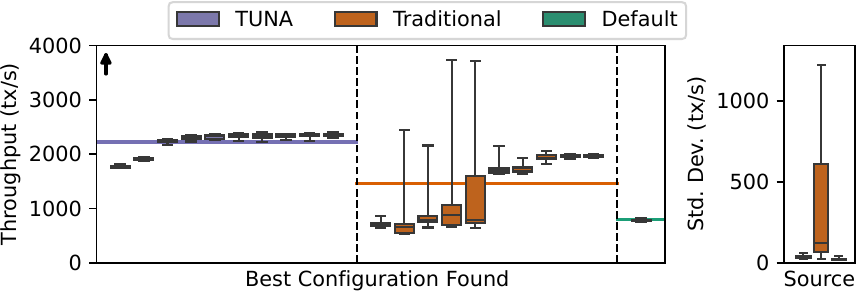}
    \Description[Noise Adjuster Model Component Analysis.]{Noise Adjuster Model Component Analysis.}
    \caption{Results of applying tuned \postgres configs running \texttt{TPC-C} with a GP optimizer. Higher is better.}
    \label{fig:eval_gp}
\end{figure}
To show the generality of \SystemName, we replace the optimizer used in both \SystemName and traditional techniques.
We tune \postgres 16.1, and run \texttt{TPC-C}, as seen in Figure~\ref{fig:eval_gp}.
On average, \SystemName achieves $53.1\%$ higher performance with $89.5\%$ lower standard deviation than traditional sampling.

\noindent\textbf{Noise Adjuster Model.}
\begin{figure}[t]
    \setlength{\abovecaptionskip}{1ex}
    \centering
    \begin{subfigure}[t]{0.374\linewidth}
    
        \includegraphics[width=\linewidth]{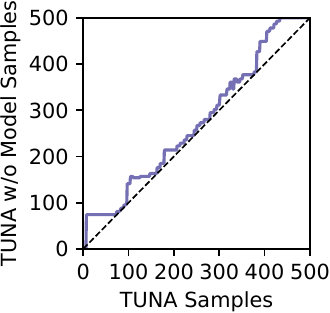}
        \caption{\added{TUNA convergence gains, with and without the noise adjuster model. Being above dashed line means TUNA is better.}} 
        \vspace{1ex}
        \label{fig:tuna_vs_nomodel}
    \end{subfigure}
    \hspace{0.03\linewidth}
    \begin{subfigure}[t]{0.57\linewidth}
        \includegraphics[width=\linewidth]{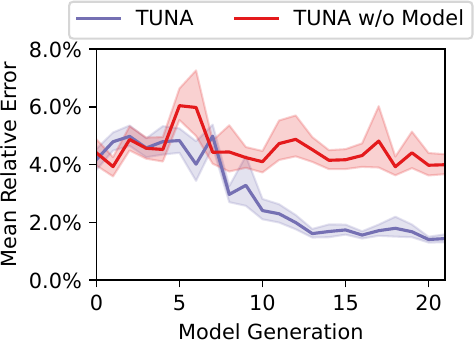}
        \caption{\changed{Comparison of relative error between reported values with and without our noise adjuster model. We show that our model can remove $67.3$\% of the noise.}}
        \vspace{1ex}
        \label{fig:model_eval}
    \end{subfigure}

    \caption{ 
    \added{
        Convergence comparisons for $100$ tuning runs on \SystemName and \SystemName without the noise adjuster model.
    }
    }
    \label{fig:model_ablation}
\end{figure}
To evaluate the performance of our noise adjuster model, we run an ablation study comparing \SystemName against itself with the noise adjuster model removed.
We tune \postgres running \texttt{epinions} with the same setup as described in \S~\ref{sec:eval_workloads}.
We run each system for $100$ tuning runs, and report the results in Figure~\ref{fig:model_ablation}.

First, we examine the convergence rates between the two systems in Figure~\ref{fig:tuna_vs_nomodel}.
This figure shows the number of iterations it takes for \SystemName to converge to the same performance as \SystemName without the noise adjuster model.
We can see immediately that the full \SystemName system is always able to converge faster than when it lacks the noise adjuster model, as the performance is always above the diagonal.
On average, the model allows \SystemName to converge $13.3\%$ faster.

This improvement in performance can be attributed to the the scores we report to the optimizer being closer to the ground truth and hence more consistent.
We compare against our ground truth value, i.e. runs with the max budget, (as described in \Section\ref{sec:noise_model}).
It is important to note that the inference step happens before the training step in our pipeline, so we do not leak any information to the noise adjuster model during this process.
The results of this experiment can be seen in Figure~\ref{fig:model_eval}.
We can see that at the beginning of the run, while the model has very few samples to learn from, it is not able to make significant improvements over a system without the model.
By the halfway mark and beyond, the error in the system without the model is $4.87\%$, whereas, on the system with the model, it is $1.99\%$, a $53.34\%$ relative reduction in error.
On average, for the duration of the entire run, the model reduces the error by $35.8\%$ including the initial runs in which the model has not seen much data.
From midpoint of the run onwards, the model reduces the error by $59.2\%$ on average.
This reduction in error allows the optimizer to have a more accurate representation of the ground truth performance for a sample.

\noindent\textbf{Impact of Outlier Detector.}
\begin{figure}[t]
    \centering
    \includegraphics[width=0.95\linewidth]{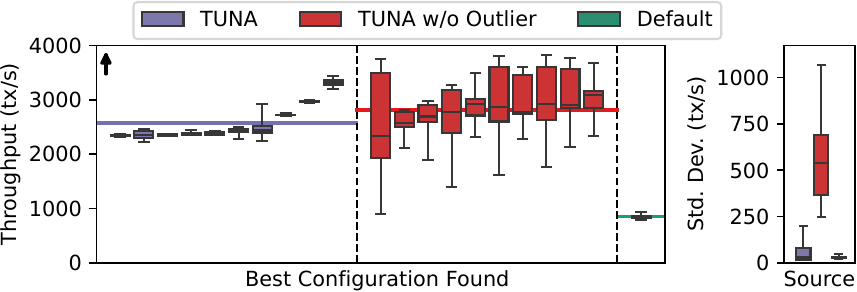}
    \Description[Noise Adjuster Model Component Analysis.]{Noise Adjuster Model Component Analysis.}
    \caption{Comparison of \SystemName with and without the Outlier Detector component running \texttt{TPC-C} on \postgres 16.1.
    \added{Although configs found using \SystemName have a slightly lower mean performance by $8.5\%$, their variability is substantially slower by $91\%$ on average.}
    }
    \label{fig:eval_outlier}
\end{figure}
To evaluate the performance of the outlier detector, we compare our full system, \SystemName, against itself with the outlier detector removed.
The result of this experiment is shown in Figure~\ref{fig:eval_outlier}.
We find that \SystemName finds performance values of $2572$ TPS, with an average standard deviation of $54.8$ TPS.
The system with the outlier detection components removed finds configurations that, on average, have $2810$ TPS and an average standard deviation of $550.8$ TPS.
Although it is not surprising that the optimizer is able to find configurations with higher performance, as this is its only objective, it is significantly limited by the fact that many of these configurations are extremely unstable.
Configurations found with \SystemName have $10.1$x lower variability than a system without these components.
From these results, we believe that this shows the necessity of \emph{unstable} configuration aware sampling.

\section{Future Work}
\label{sec:future_work}
We see four main areas of future work to further improve this sampling methodology.
First, we did not constrain how much our noise model (\S~\ref{sec:noise_model}) can adjust the result.
While we did not experience any issues during evaluation, or testing, it is foreseeable that there exists a pattern of data that could cause unpredictable results from the model.
For a production setting, it may be wise to include guardrails limiting how much the model can adjust the score for the optimizer.
Building these guardrails would, however, require additional knowledge about the variability of the system.

An alternative approach to the outlier detector would be to have a scaling penalty based on the relative range seen, rather than a penalty for crossing a threshold.
This is something that has been studied in economics~\cite{MarkowitzMeanVariance, MeanSemiVariance, MeanAbsoluteDeviation, MeanVarianceSkewModel, MeanSemiVariance2}, but for the context of tuning, a hyperparameter for penalty is required.
This is something that we wished to avoid.

Our noise adjuster model only utilizes data for predictions from within a run which can delay its effectiveness.
Alternatively, we could transfer data from prior runs to warm-start the model.
This has technical challenges, particularly when using a different set of machines or workloads for warm-start information.
This is in many ways similar to the transfer of the surrogate model in the context of autotuning.

Finally, we do not address burstable or serverless nodes, in large part because it is difficult to differentiate between an unstable configuration and a credit depletion.
We ignore this as without platform introspection it is very difficult to determine the root cause.
Additionally, a system that is targeting burstable VMs should generate a configuration that can perform well for both a VM with and without credits.


\section{Conclusion}
\label{sec:conclusion}
In this paper, we explored ideas around the impacts of performance noise on cloud system \autotuning, and discovered two main problems.
If one does not account for performance variability during tuning, convergence rates to an optimal configuration are slower and more costly, and configurations may be unstable during deployment.
We further motivate our system by investigating the magnitude of noise in the cloud, and that while some components have virtually no noise, others have much higher noise than bare metal components.

We solve these problems through three main mechanisms: a multi-fidelity sampling process, an outlier detector, and a simple model built from component-level metrics to mitigate noise during sampling.
We then integrate this into a state-of-the-art tuning setup and evaluate it across 6 workloads, 3 systems, 2 deployment regions, and 2 hardware setups, finding that \SystemName reduces variability and improves the performance of tuned systems.  

\added{
\section*{Acknowledgments}
\label{sec:acknowledgment}
We would like to thank our anonymous reviewers and our shepherd, Thomas Pasquier.
We would also like to thank the Microsoft Gray Systems Lab for providing Azure cloud credits and for early feedback from several members, including Carlo Curino, Yuanyuan Tian, and Andreas Mueller.
This material is based upon work supported by the National Science Foundation under Grant No. 2326576.

}

\bibliographystyle{ACM-Reference-Format} 
\bibliography{references}


\begin{thebibliography}{104}


\ifx \showCODEN    \undefined \def \showCODEN     #1{\unskip}     \fi
\ifx \showDOI      \undefined \def \showDOI       #1{#1}\fi
\ifx \showISBNx    \undefined \def \showISBNx     #1{\unskip}     \fi
\ifx \showISBNxiii \undefined \def \showISBNxiii  #1{\unskip}     \fi
\ifx \showISSN     \undefined \def \showISSN      #1{\unskip}     \fi
\ifx \showLCCN     \undefined \def \showLCCN      #1{\unskip}     \fi
\ifx \shownote     \undefined \def \shownote      #1{#1}          \fi
\ifx \showarticletitle \undefined \def \showarticletitle #1{#1}   \fi
\ifx \showURL      \undefined \def \showURL       {\relax}        \fi
\providecommand\bibfield[2]{#2}
\providecommand\bibinfo[2]{#2}
\providecommand\natexlab[1]{#1}
\providecommand\showeprint[2][]{arXiv:#2}

\bibitem[Abedi and Brecht(2017a)]%
        {testordering}
\bibfield{author}{\bibinfo{person}{Ali Abedi} {and} \bibinfo{person}{Tim Brecht}.} \bibinfo{year}{2017}\natexlab{a}.
\newblock \showarticletitle{Conducting Repeatable Experiments in Highly Variable Cloud Computing Environments}. In \bibinfo{booktitle}{\emph{Proceedings of the 8th ACM/SPEC on International Conference on Performance Engineering}} (L'Aquila, Italy) \emph{(\bibinfo{series}{ICPE '17})}. \bibinfo{publisher}{Association for Computing Machinery}, \bibinfo{address}{New York, NY, USA}, \bibinfo{pages}{287–292}.
\newblock
\showISBNx{9781450344043}
\urldef\tempurl%
\url{https://doi.org/10.1145/3030207.3030229}
\showDOI{\tempurl}


\bibitem[Abedi and Brecht(2017b)]%
        {CloudPerfVariability2}
\bibfield{author}{\bibinfo{person}{Ali Abedi} {and} \bibinfo{person}{Tim Brecht}.} \bibinfo{year}{2017}\natexlab{b}.
\newblock \showarticletitle{Conducting Repeatable Experiments in Highly Variable Cloud Computing Environments}. In \bibinfo{booktitle}{\emph{Proceedings of the 8th ACM/SPEC on International Conference on Performance Engineering}} (L'Aquila, Italy) \emph{(\bibinfo{series}{ICPE '17})}. \bibinfo{publisher}{Association for Computing Machinery}, \bibinfo{address}{New York, NY, USA}, \bibinfo{pages}{287–292}.
\newblock
\showISBNx{9781450344043}
\urldef\tempurl%
\url{https://doi.org/10.1145/3030207.3030229}
\showDOI{\tempurl}


\bibitem[Agrawal et~al\mbox{.}(2005)]%
        {sql_server_tuning}
\bibfield{author}{\bibinfo{person}{Sanjay Agrawal}, \bibinfo{person}{Surajit Chaudhuri}, \bibinfo{person}{Lubor Kollar}, \bibinfo{person}{Arun Marathe}, \bibinfo{person}{Vivek Narasayya}, {and} \bibinfo{person}{Manoj Syamala}.} \bibinfo{year}{2005}\natexlab{}.
\newblock \showarticletitle{Database tuning advisor for microsoft SQL server 2005: demo}. In \bibinfo{booktitle}{\emph{Proceedings of the 2005 ACM SIGMOD International Conference on Management of Data}} (Baltimore, Maryland) \emph{(\bibinfo{series}{SIGMOD '05})}. \bibinfo{publisher}{Association for Computing Machinery}, \bibinfo{address}{New York, NY, USA}, \bibinfo{pages}{930–932}.
\newblock
\showISBNx{1595930604}
\urldef\tempurl%
\url{https://doi.org/10.1145/1066157.1066292}
\showDOI{\tempurl}


\bibitem[Amaral et~al\mbox{.}(2018)]%
        {SPECCPU2017}
\bibfield{author}{\bibinfo{person}{Jose~Nelson Amaral}, \bibinfo{person}{Edson Borin}, \bibinfo{person}{Dylan~R. Ashley}, \bibinfo{person}{Caian Benedicto}, \bibinfo{person}{Elliot Colp}, \bibinfo{person}{Joao Henrique~Stange Hoffmam}, \bibinfo{person}{Marcus Karpoff}, \bibinfo{person}{Erick Ochoa}, \bibinfo{person}{Morgan Redshaw}, {and} \bibinfo{person}{Raphael~Ernani Rodrigues}.} \bibinfo{year}{2018}\natexlab{}.
\newblock \showarticletitle{The Alberta Workloads for the SPEC CPU 2017 Benchmark Suite}. In \bibinfo{booktitle}{\emph{2018 IEEE International Symposium on Performance Analysis of Systems and Software (ISPASS)}}. \bibinfo{pages}{159--168}.
\newblock
\urldef\tempurl%
\url{https://doi.org/10.1109/ISPASS.2018.00029}
\showDOI{\tempurl}


\bibitem[Amazon(2024a)]%
        {AWSDisks}
Amazon \bibinfo{year}{2024}\natexlab{a}.
\newblock \bibinfo{title}{Amazon EBS volume types}.
\newblock
\newblock
\urldef\tempurl%
\url{https://docs.aws.amazon.com/ebs/latest/userguide/ebs-volume-types.html}
\showURL{%
\tempurl}


\bibitem[Amazon(2024b)]%
        {EC2Offerings}
Amazon \bibinfo{year}{2024}\natexlab{b}.
\newblock \bibinfo{title}{Amazon EC2 Instance types}.
\newblock
\newblock
\urldef\tempurl%
\url{https://aws.amazon.com/ec2/instance-types/}
\showURL{%
\tempurl}


\bibitem[Asyabi et~al\mbox{.}(2018)]%
        {ppXen}
\bibfield{author}{\bibinfo{person}{Esmail Asyabi}, \bibinfo{person}{Mohsen Sharifi}, {and} \bibinfo{person}{Azer Bestavros}.} \bibinfo{year}{2018}\natexlab{}.
\newblock \showarticletitle{ppXen: A hypervisor CPU scheduler for mitigating performance variability in virtualized clouds}.
\newblock \bibinfo{journal}{\emph{Future Generation Computer Systems}}  \bibinfo{volume}{83} (\bibinfo{year}{2018}), \bibinfo{pages}{75--84}.
\newblock
\showISSN{0167-739X}
\urldef\tempurl%
\url{https://doi.org/10.1016/j.future.2018.01.015}
\showDOI{\tempurl}


\bibitem[Atla et~al\mbox{.}(2011)]%
        {MLNoiseSensitivity}
\bibfield{author}{\bibinfo{person}{Abhinav Atla}, \bibinfo{person}{Rahul Tada}, \bibinfo{person}{Victor Sheng}, {and} \bibinfo{person}{Naveen Singireddy}.} \bibinfo{year}{2011}\natexlab{}.
\newblock \showarticletitle{Sensitivity of different machine learning algorithms to noise}.
\newblock \bibinfo{journal}{\emph{J. Comput. Sci. Coll.}} \bibinfo{volume}{26}, \bibinfo{number}{5} (\bibinfo{date}{may} \bibinfo{year}{2011}), \bibinfo{pages}{96–103}.
\newblock
\showISSN{1937-4771}
\urldef\tempurl%
\url{https://dl.acm.org/doi/abs/10.5555/1961574.1961594}
\showURL{%
\tempurl}


\bibitem[Axboe and Fum(2021)]%
        {fio}
\bibfield{author}{\bibinfo{person}{Jens Axboe} {and} \bibinfo{person}{Vincent Fum}.} \bibinfo{year}{2021}\natexlab{}.
\newblock \bibinfo{title}{Fio}.
\newblock
\newblock
\urldef\tempurl%
\url{https://github.com/axboe/fio}
\showURL{%
\tempurl}


\bibitem[Balakrishnan et~al\mbox{.}(2005)]%
        {CpuCorePerfVariance}
\bibfield{author}{\bibinfo{person}{S. Balakrishnan}, \bibinfo{person}{Ravi Rajwar}, \bibinfo{person}{M. Upton}, {and} \bibinfo{person}{K. Lai}.} \bibinfo{year}{2005}\natexlab{}.
\newblock \showarticletitle{The impact of performance asymmetry in emerging multicore architectures}. In \bibinfo{booktitle}{\emph{32nd International Symposium on Computer Architecture (ISCA'05)}}. \bibinfo{publisher}{IEEE}, \bibinfo{address}{Madison, WI}, \bibinfo{pages}{506--517}.
\newblock
\urldef\tempurl%
\url{https://doi.org/10.1109/ISCA.2005.51}
\showDOI{\tempurl}


\bibitem[Balandat et~al\mbox{.}(2020)]%
        {botorch}
\bibfield{author}{\bibinfo{person}{Maximilian Balandat}, \bibinfo{person}{Brian Karrer}, \bibinfo{person}{Daniel Jiang}, \bibinfo{person}{Samuel Daulton}, \bibinfo{person}{Ben Letham}, \bibinfo{person}{Andrew~G Wilson}, {and} \bibinfo{person}{Eytan Bakshy}.} \bibinfo{year}{2020}\natexlab{}.
\newblock \showarticletitle{BoTorch: A Framework for Efficient Monte-Carlo Bayesian Optimization}. In \bibinfo{booktitle}{\emph{Advances in Neural Information Processing Systems}}, \bibfield{editor}{\bibinfo{person}{H.~Larochelle}, \bibinfo{person}{M.~Ranzato}, \bibinfo{person}{R.~Hadsell}, \bibinfo{person}{M.F. Balcan}, {and} \bibinfo{person}{H.~Lin}} (Eds.), Vol.~\bibinfo{volume}{33}. \bibinfo{publisher}{Curran Associates, Inc.}, \bibinfo{pages}{21524--21538}.
\newblock
\urldef\tempurl%
\url{https://proceedings.neurips.cc/paper_files/paper/2020/file/f5b1b89d98b7286673128a5fb112cb9a-Paper.pdf}
\showURL{%
\tempurl}


\bibitem[Cao et~al\mbox{.}(2018)]%
        {storagetuning}
\bibfield{author}{\bibinfo{person}{Zhen Cao}, \bibinfo{person}{Vasily Tarasov}, \bibinfo{person}{Sachin Tiwari}, {and} \bibinfo{person}{Erez Zadok}.} \bibinfo{year}{2018}\natexlab{}.
\newblock \showarticletitle{Towards Better Understanding of Black-box {Auto-Tuning}: A Comparative Analysis for Storage Systems}. In \bibinfo{booktitle}{\emph{2018 USENIX Annual Technical Conference (USENIX ATC 18)}}. \bibinfo{publisher}{USENIX Association}, \bibinfo{address}{Boston, MA}, \bibinfo{pages}{893--907}.
\newblock
\showISBNx{978-1-939133-01-4}
\urldef\tempurl%
\url{https://dl.acm.org/doi/10.5555/3277355.3277441}
\showURL{%
\tempurl}


\bibitem[Casale and Tribastone(2013)]%
        {outlierdetection2}
\bibfield{author}{\bibinfo{person}{Giuliano Casale} {and} \bibinfo{person}{Mirco Tribastone}.} \bibinfo{year}{2013}\natexlab{}.
\newblock \showarticletitle{Modelling exogenous variability in cloud deployments}.
\newblock \bibinfo{journal}{\emph{SIGMETRICS Perform. Eval. Rev.}} \bibinfo{volume}{40}, \bibinfo{number}{4} (\bibinfo{date}{April} \bibinfo{year}{2013}), \bibinfo{pages}{73–82}.
\newblock
\showISSN{0163-5999}
\urldef\tempurl%
\url{https://doi.org/10.1145/2479942.2479951}
\showDOI{\tempurl}


\bibitem[Chang et~al\mbox{.}(2009)]%
        {MeanSemiVariance}
\bibfield{author}{\bibinfo{person}{Tun-Jen Chang}, \bibinfo{person}{Sang-Chin Yang}, {and} \bibinfo{person}{Kuang-Jung Chang}.} \bibinfo{year}{2009}\natexlab{}.
\newblock \showarticletitle{Portfolio optimization problems in different risk measures using genetic algorithm}.
\newblock \bibinfo{journal}{\emph{Expert Systems with Applications}} \bibinfo{volume}{36}, \bibinfo{number}{7} (\bibinfo{year}{2009}), \bibinfo{pages}{10529--10537}.
\newblock
\showISSN{0957-4174}
\urldef\tempurl%
\url{https://doi.org/10.1016/j.eswa.2009.02.062}
\showDOI{\tempurl}


\bibitem[Chatterjee et~al\mbox{.}(2021)]%
        {cosine_kv}
\bibfield{author}{\bibinfo{person}{Subarna Chatterjee}, \bibinfo{person}{Meena Jagadeesan}, \bibinfo{person}{Wilson Qin}, {and} \bibinfo{person}{Stratos Idreos}.} \bibinfo{year}{2021}\natexlab{}.
\newblock \showarticletitle{Cosine: a cloud-cost optimized self-designing key-value storage engine}.
\newblock \bibinfo{journal}{\emph{Proc. VLDB Endow.}} \bibinfo{volume}{15}, \bibinfo{number}{1} (\bibinfo{date}{Sept.} \bibinfo{year}{2021}), \bibinfo{pages}{112–126}.
\newblock
\showISSN{2150-8097}
\urldef\tempurl%
\url{https://doi.org/10.14778/3485450.3485461}
\showDOI{\tempurl}


\bibitem[Chen et~al\mbox{.}(2019)]%
        {PARTIES}
\bibfield{author}{\bibinfo{person}{Shuang Chen}, \bibinfo{person}{Christina Delimitrou}, {and} \bibinfo{person}{Jos\'{e}~F. Mart\'{\i}nez}.} \bibinfo{year}{2019}\natexlab{}.
\newblock \showarticletitle{PARTIES: QoS-Aware Resource Partitioning for Multiple Interactive Services}. In \bibinfo{booktitle}{\emph{Proceedings of the Twenty-Fourth International Conference on Architectural Support for Programming Languages and Operating Systems}} (Providence, RI, USA) \emph{(\bibinfo{series}{ASPLOS '19})}. \bibinfo{publisher}{Association for Computing Machinery}, \bibinfo{address}{New York, NY, USA}, \bibinfo{pages}{107–120}.
\newblock
\showISBNx{9781450362405}
\urldef\tempurl%
\url{https://doi.org/10.1145/3297858.3304005}
\showDOI{\tempurl}


\bibitem[Cooper et~al\mbox{.}(2010)]%
        {YCSB}
\bibfield{author}{\bibinfo{person}{Brian~F. Cooper}, \bibinfo{person}{Adam Silberstein}, \bibinfo{person}{Erwin Tam}, \bibinfo{person}{Raghu Ramakrishnan}, {and} \bibinfo{person}{Russell Sears}.} \bibinfo{year}{2010}\natexlab{}.
\newblock \showarticletitle{Benchmarking Cloud Serving Systems with YCSB}. In \bibinfo{booktitle}{\emph{Proceedings of the 1st ACM Symposium on Cloud Computing}} (Indianapolis, Indiana, USA) \emph{(\bibinfo{series}{SoCC '10})}. \bibinfo{publisher}{Association for Computing Machinery}, \bibinfo{address}{New York, NY, USA}, \bibinfo{pages}{143–154}.
\newblock
\showISBNx{9781450300360}
\urldef\tempurl%
\url{https://doi.org/10.1145/1807128.1807152}
\showDOI{\tempurl}


\bibitem[Cortez et~al\mbox{.}(2017)]%
        {ResourceCentral}
\bibfield{author}{\bibinfo{person}{Eli Cortez}, \bibinfo{person}{Anand Bonde}, \bibinfo{person}{Alexandre Muzio}, \bibinfo{person}{Mark Russinovich}, \bibinfo{person}{Marcus Fontoura}, {and} \bibinfo{person}{Ricardo Bianchini}.} \bibinfo{year}{2017}\natexlab{}.
\newblock \showarticletitle{Resource Central: Understanding and Predicting Workloads for Improved Resource Management in Large Cloud Platforms}. In \bibinfo{booktitle}{\emph{Proceedings of the 26th Symposium on Operating Systems Principles}} (Shanghai, China) \emph{(\bibinfo{series}{SOSP '17})}. \bibinfo{publisher}{Association for Computing Machinery}, \bibinfo{address}{New York, NY, USA}, \bibinfo{pages}{153–167}.
\newblock
\showISBNx{9781450350853}
\urldef\tempurl%
\url{https://doi.org/10.1145/3132747.3132772}
\showDOI{\tempurl}


\bibitem[De~Sensi et~al\mbox{.}(2022)]%
        {NoiseInTheClouds}
\bibfield{author}{\bibinfo{person}{Daniele De~Sensi}, \bibinfo{person}{Tiziano De~Matteis}, \bibinfo{person}{Konstantin Taranov}, \bibinfo{person}{Salvatore Di~Girolamo}, \bibinfo{person}{Tobias Rahn}, {and} \bibinfo{person}{Torsten Hoefler}.} \bibinfo{year}{2022}\natexlab{}.
\newblock \showarticletitle{Noise in the Clouds: Influence of Network Performance Variability on Application Scalability}.
\newblock \bibinfo{journal}{\emph{Proc. ACM Meas. Anal. Comput. Syst.}} \bibinfo{volume}{6}, \bibinfo{number}{3}, Article \bibinfo{articleno}{49} (\bibinfo{date}{Dec.} \bibinfo{year}{2022}), \bibinfo{numpages}{27}~pages.
\newblock
\urldef\tempurl%
\url{https://doi.org/10.1145/3570609}
\showDOI{\tempurl}


\bibitem[Dean and Ghemawat(2004)]%
        {MapReduce}
\bibfield{author}{\bibinfo{person}{Jeffrey Dean} {and} \bibinfo{person}{Sanjay Ghemawat}.} \bibinfo{year}{2004}\natexlab{}.
\newblock \showarticletitle{MapReduce: Simplified Data Processing on Large Clusters}. In \bibinfo{booktitle}{\emph{OSDI'04: Sixth Symposium on Operating System Design and Implementation}}. \bibinfo{address}{San Francisco, CA}, \bibinfo{pages}{137--150}.
\newblock
\urldef\tempurl%
\url{https://doi.org/10.1145/1327452.1327492}
\showURL{%
\tempurl}


\bibitem[Difallah et~al\mbox{.}(2013)]%
        {Benchbase}
\bibfield{author}{\bibinfo{person}{Djellel~Eddine Difallah}, \bibinfo{person}{Andrew Pavlo}, \bibinfo{person}{Carlo Curino}, {and} \bibinfo{person}{Philippe Cudr{\'e}-Mauroux}.} \bibinfo{year}{2013}\natexlab{}.
\newblock \showarticletitle{OLTP-Bench: An Extensible Testbed for Benchmarking Relational Databases}.
\newblock \bibinfo{journal}{\emph{PVLDB}} \bibinfo{volume}{7}, \bibinfo{number}{4} (\bibinfo{year}{2013}), \bibinfo{pages}{277--288}.
\newblock
\urldef\tempurl%
\url{http://www.vldb.org/pvldb/vol7/p277-difallah.pdf}
\showURL{%
\tempurl}


\bibitem[Duan et~al\mbox{.}(2009)]%
        {ituned}
\bibfield{author}{\bibinfo{person}{Songyun Duan}, \bibinfo{person}{Vamsidhar Thummala}, {and} \bibinfo{person}{Shivnath Babu}.} \bibinfo{year}{2009}\natexlab{}.
\newblock \showarticletitle{Tuning database configuration parameters with iTuned}.
\newblock \bibinfo{journal}{\emph{Proc. VLDB Endow.}} \bibinfo{volume}{2}, \bibinfo{number}{1} (\bibinfo{date}{Aug.} \bibinfo{year}{2009}), \bibinfo{pages}{1246–1257}.
\newblock
\showISSN{2150-8097}
\urldef\tempurl%
\url{https://doi.org/10.14778/1687627.1687767}
\showDOI{\tempurl}


\bibitem[Duplyakin et~al\mbox{.}(2019)]%
        {Cloudlab}
\bibfield{author}{\bibinfo{person}{Dmitry Duplyakin}, \bibinfo{person}{Robert Ricci}, \bibinfo{person}{Aleksander Maricq}, \bibinfo{person}{Gary Wong}, \bibinfo{person}{Jonathon Duerig}, \bibinfo{person}{Eric Eide}, \bibinfo{person}{Leigh Stoller}, \bibinfo{person}{Mike Hibler}, \bibinfo{person}{David Johnson}, \bibinfo{person}{Kirk Webb}, \bibinfo{person}{Aditya Akella}, \bibinfo{person}{Kuangching Wang}, \bibinfo{person}{Glenn Ricart}, \bibinfo{person}{Larry Landweber}, \bibinfo{person}{Chip Elliott}, \bibinfo{person}{Michael Zink}, \bibinfo{person}{Emmanuel Cecchet}, \bibinfo{person}{Snigdhaswin Kar}, {and} \bibinfo{person}{Prabodh Mishra}.} \bibinfo{year}{2019}\natexlab{}.
\newblock \showarticletitle{The Design and Operation of {CloudLab}}. In \bibinfo{booktitle}{\emph{2019 USENIX Annual Technical Conference (USENIX ATC 19)}}. \bibinfo{publisher}{USENIX Association}, \bibinfo{address}{Renton, WA}, \bibinfo{pages}{1--14}.
\newblock
\showISBNx{978-1-939133-03-8}
\urldef\tempurl%
\url{https://dl.acm.org/doi/10.5555/3358807.3358809}
\showURL{%
\tempurl}


\bibitem[Ericson et~al\mbox{.}(2017)]%
        {CloudPerfVariability}
\bibfield{author}{\bibinfo{person}{Jamie Ericson}, \bibinfo{person}{Masoud Mohammadian}, {and} \bibinfo{person}{Fabiana Santana}.} \bibinfo{year}{2017}\natexlab{}.
\newblock \showarticletitle{Analysis of Performance Variability in Public Cloud Computing}. In \bibinfo{booktitle}{\emph{2017 IEEE International Conference on Information Reuse and Integration (IRI)}}. \bibinfo{publisher}{IEEE}, \bibinfo{address}{San Diego, CA, USA}, \bibinfo{pages}{308--314}.
\newblock
\urldef\tempurl%
\url{https://doi.org/10.1109/IRI.2017.47}
\showDOI{\tempurl}


\bibitem[Farley et~al\mbox{.}(2012)]%
        {Moreforyourmoney}
\bibfield{author}{\bibinfo{person}{Benjamin Farley}, \bibinfo{person}{Ari Juels}, \bibinfo{person}{Venkatanathan Varadarajan}, \bibinfo{person}{Thomas Ristenpart}, \bibinfo{person}{Kevin~D. Bowers}, {and} \bibinfo{person}{Michael~M. Swift}.} \bibinfo{year}{2012}\natexlab{}.
\newblock \showarticletitle{More for Your Money: Exploiting Performance Heterogeneity in Public Clouds}. In \bibinfo{booktitle}{\emph{Proceedings of the Third ACM Symposium on Cloud Computing}} (San Jose, California) \emph{(\bibinfo{series}{SoCC '12})}. \bibinfo{publisher}{Association for Computing Machinery}, \bibinfo{address}{New York, NY, USA}, Article \bibinfo{articleno}{20}, \bibinfo{numpages}{14}~pages.
\newblock
\showISBNx{9781450317610}
\urldef\tempurl%
\url{https://doi.org/10.1145/2391229.2391249}
\showDOI{\tempurl}


\bibitem[Fekry et~al\mbox{.}(2020)]%
        {tuneornot}
\bibfield{author}{\bibinfo{person}{Ayat Fekry}, \bibinfo{person}{Lucian Carata}, \bibinfo{person}{Thomas Pasquier}, \bibinfo{person}{Andrew Rice}, {and} \bibinfo{person}{Andy Hopper}.} \bibinfo{year}{2020}\natexlab{}.
\newblock \showarticletitle{To Tune or Not to Tune? In Search of Optimal Configurations for Data Analytics}. In \bibinfo{booktitle}{\emph{Proceedings of the 26th ACM SIGKDD International Conference on Knowledge Discovery \& Data Mining}} (Virtual Event, CA, USA) \emph{(\bibinfo{series}{KDD '20})}. \bibinfo{publisher}{Association for Computing Machinery}, \bibinfo{address}{New York, NY, USA}, \bibinfo{pages}{2494–2504}.
\newblock
\showISBNx{9781450379984}
\urldef\tempurl%
\url{https://doi.org/10.1145/3394486.3403299}
\showDOI{\tempurl}


\bibitem[Figiela et~al\mbox{.}(2018)]%
        {PerformanceEvalCloudFunctions}
\bibfield{author}{\bibinfo{person}{Kamil Figiela}, \bibinfo{person}{Adam Gajek}, \bibinfo{person}{Adam Zima}, \bibinfo{person}{Beata Obrok}, {and} \bibinfo{person}{Maciej Malawski}.} \bibinfo{year}{2018}\natexlab{}.
\newblock \showarticletitle{Performance evaluation of heterogeneous cloud functions}.
\newblock \bibinfo{journal}{\emph{Concurrency and Computation: Practice and Experience}} \bibinfo{volume}{30}, \bibinfo{number}{23} (\bibinfo{year}{2018}).
\newblock
\urldef\tempurl%
\url{https://doi.org/10.1002/cpe.4792}
\showURL{%
\tempurl}


\bibitem[Frazier(2018)]%
        {BOWithNoise}
\bibfield{author}{\bibinfo{person}{Peter~I Frazier}.} \bibinfo{year}{2018}\natexlab{}.
\newblock \showarticletitle{A tutorial on Bayesian optimization}.
\newblock \bibinfo{journal}{\emph{arXiv preprint arXiv:1807.02811}} (\bibinfo{year}{2018}).
\newblock
\urldef\tempurl%
\url{https://doi.org/10.48550/arXiv.1807.02811}
\showURL{%
\tempurl}


\bibitem[Geelnard(2023)]%
        {osbench}
\bibfield{author}{\bibinfo{person}{Marcus Geelnard}.} \bibinfo{year}{2023}\natexlab{}.
\newblock \bibinfo{title}{osbench}.
\newblock
\newblock
\urldef\tempurl%
\url{https://gitlab.com/mbitsnbites/osbench}
\showURL{%
\tempurl}


\bibitem[Google(2024a)]%
        {GCPVendorBenchmarks}
Google \bibinfo{year}{2024}\natexlab{a}.
\newblock \bibinfo{title}{CoreMark scores of VMs by family}.
\newblock
\newblock
\urldef\tempurl%
\url{https://cloud.google.com/compute/docs/coremark-scores-of-vm-instances}
\showURL{%
\tempurl}


\bibitem[Google(2024b)]%
        {GCPDisks}
Google \bibinfo{year}{2024}\natexlab{b}.
\newblock \bibinfo{title}{Storage options}.
\newblock
\newblock
\urldef\tempurl%
\url{https://cloud.google.com/compute/docs/disks}
\showURL{%
\tempurl}


\bibitem[Gray(1986)]%
        {OriginalGray}
\bibfield{author}{\bibinfo{person}{Jim Gray}.} \bibinfo{year}{1986}\natexlab{}.
\newblock \showarticletitle{Why do computers stop and what can be done about it?}. In \bibinfo{booktitle}{\emph{Symposium on reliability in distributed software and database systems}}. Los Angeles, CA, USA, \bibinfo{pages}{3--12}.
\newblock
\urldef\tempurl%
\url{https://doi.org/10.1109/MC.1985.1662717}
\showURL{%
\tempurl}


\bibitem[Group(2022)]%
        {pgbench}
\bibfield{author}{\bibinfo{person}{PostgreSQL Global~Development Group}.} \bibinfo{year}{2022}\natexlab{}.
\newblock \bibinfo{title}{pgbench}.
\newblock
\newblock
\urldef\tempurl%
\url{https://www.postgresql.org/docs/15/pgbench.html}
\showURL{%
\tempurl}


\bibitem[Gupta et~al\mbox{.}(2016)]%
        {HPCInterference}
\bibfield{author}{\bibinfo{person}{Abhishek Gupta}, \bibinfo{person}{Paolo Faraboschi}, \bibinfo{person}{Filippo Gioachin}, \bibinfo{person}{Laxmikant~V. Kale}, \bibinfo{person}{Richard Kaufmann}, \bibinfo{person}{Bu-Sung Lee}, \bibinfo{person}{Verdi March}, \bibinfo{person}{Dejan Milojicic}, {and} \bibinfo{person}{Chun~Hui Suen}.} \bibinfo{year}{2016}\natexlab{}.
\newblock \showarticletitle{Evaluating and Improving the Performance and Scheduling of HPC Applications in Cloud}.
\newblock \bibinfo{journal}{\emph{IEEE Transactions on Cloud Computing}} \bibinfo{volume}{4}, \bibinfo{number}{3} (\bibinfo{year}{2016}), \bibinfo{pages}{307--321}.
\newblock
\urldef\tempurl%
\url{https://doi.org/10.1109/TCC.2014.2339858}
\showDOI{\tempurl}


\bibitem[Huang et~al\mbox{.}(2017)]%
        {GrayFailure1}
\bibfield{author}{\bibinfo{person}{Peng Huang}, \bibinfo{person}{Chuanxiong Guo}, \bibinfo{person}{Lidong Zhou}, \bibinfo{person}{Jacob~R. Lorch}, \bibinfo{person}{Yingnong Dang}, \bibinfo{person}{Murali Chintalapati}, {and} \bibinfo{person}{Randolph Yao}.} \bibinfo{year}{2017}\natexlab{}.
\newblock \showarticletitle{Gray Failure: The Achilles' Heel of Cloud-Scale Systems}. In \bibinfo{booktitle}{\emph{Proceedings of the 16th Workshop on Hot Topics in Operating Systems}} (Whistler, BC, Canada) \emph{(\bibinfo{series}{HotOS '17})}. \bibinfo{publisher}{Association for Computing Machinery}, \bibinfo{address}{New York, NY, USA}, \bibinfo{pages}{150–155}.
\newblock
\showISBNx{9781450350686}
\urldef\tempurl%
\url{https://doi.org/10.1145/3102980.3103005}
\showDOI{\tempurl}


\bibitem[Hutter et~al\mbox{.}(2011)]%
        {SMAC}
\bibfield{author}{\bibinfo{person}{Frank Hutter}, \bibinfo{person}{Holger~H. Hoos}, {and} \bibinfo{person}{Kevin Leyton-Brown}.} \bibinfo{year}{2011}\natexlab{}.
\newblock \showarticletitle{Sequential Model-Based Optimization for General Algorithm Configuration}. In \bibinfo{booktitle}{\emph{Learning and Intelligent Optimization}}, \bibfield{editor}{\bibinfo{person}{Carlos A.~Coello Coello}} (Ed.). \bibinfo{publisher}{Springer Berlin Heidelberg}, \bibinfo{address}{Berlin, Heidelberg}, \bibinfo{pages}{507--523}.
\newblock
\showISBNx{978-3-642-25566-3}
\newblock
\shownote{\url{https://doi.org/10.1007/978-3-642-25566-3_40}}.


\bibitem[Iosup et~al\mbox{.}(2011)]%
        {LongCloudPerfProd}
\bibfield{author}{\bibinfo{person}{Alexandru Iosup}, \bibinfo{person}{Nezih Yigitbasi}, {and} \bibinfo{person}{Dick Epema}.} \bibinfo{year}{2011}\natexlab{}.
\newblock \showarticletitle{On the Performance Variability of Production Cloud Services}. In \bibinfo{booktitle}{\emph{2011 11th IEEE/ACM International Symposium on Cluster, Cloud and Grid Computing}}. \bibinfo{publisher}{IEEE}, \bibinfo{address}{Newport Beach, CA, USA}, \bibinfo{pages}{104--113}.
\newblock
\urldef\tempurl%
\url{https://doi.org/10.1109/CCGrid.2011.22}
\showDOI{\tempurl}


\bibitem[Jamieson and Talwalkar(2015)]%
        {SuccessiveHalving}
\bibfield{author}{\bibinfo{person}{Kevin Jamieson} {and} \bibinfo{person}{Ameet Talwalkar}.} \bibinfo{year}{2015}\natexlab{}.
\newblock \bibinfo{title}{Non-stochastic Best Arm Identification and Hyperparameter Optimization}.
\newblock
\newblock
\showeprint[arxiv]{1502.07943}~[cs.LG]
\urldef\tempurl%
\url{https://doi.org/10.48550/arXiv.1502.07943}
\showURL{%
\tempurl}


\bibitem[Jia et~al\mbox{.}(2023)]%
        {D_Series}
\bibfield{author}{\bibinfo{person}{Andy Jia}, \bibinfo{person}{Micah McKittrick}, \bibinfo{person}{Styli Tsinaroglou}, {and} \bibinfo{person}{Joel Pelley}.} \bibinfo{year}{2023}\natexlab{}.
\newblock \bibinfo{title}{DV5 and DSV5-Series - Azure Virtual Machines}.
\newblock
\newblock
\urldef\tempurl%
\url{https://learn.microsoft.com/en-us/azure/virtual-machines/dv5-dsv5-series}
\showURL{%
\tempurl}


\bibitem[Joshi(2012a)]%
        {hadoop_tuning_no_ml}
\bibfield{author}{\bibinfo{person}{Shrinivas~B. Joshi}.} \bibinfo{year}{2012}\natexlab{a}.
\newblock \showarticletitle{Apache hadoop performance-tuning methodologies and best practices}. In \bibinfo{booktitle}{\emph{Proceedings of the 3rd ACM/SPEC International Conference on Performance Engineering}} (Boston, Massachusetts, USA) \emph{(\bibinfo{series}{ICPE '12})}. \bibinfo{publisher}{Association for Computing Machinery}, \bibinfo{address}{New York, NY, USA}, \bibinfo{pages}{241–242}.
\newblock
\showISBNx{9781450312028}
\urldef\tempurl%
\url{https://doi.org/10.1145/2188286.2188323}
\showDOI{\tempurl}


\bibitem[Joshi(2012b)]%
        {memory_tuning_no_ml}
\bibfield{author}{\bibinfo{person}{Shrinivas~B. Joshi}.} \bibinfo{year}{2012}\natexlab{b}.
\newblock \showarticletitle{Apache hadoop performance-tuning methodologies and best practices}. In \bibinfo{booktitle}{\emph{Proceedings of the 3rd ACM/SPEC International Conference on Performance Engineering}} (Boston, Massachusetts, USA) \emph{(\bibinfo{series}{ICPE '12})}. \bibinfo{publisher}{Association for Computing Machinery}, \bibinfo{address}{New York, NY, USA}, \bibinfo{pages}{241–242}.
\newblock
\showISBNx{9781450312028}
\urldef\tempurl%
\url{https://doi.org/10.1145/2188286.2188323}
\showDOI{\tempurl}


\bibitem[Kanellis et~al\mbox{.}(2022)]%
        {Llamatune}
\bibfield{author}{\bibinfo{person}{Konstantinos Kanellis}, \bibinfo{person}{Cong Ding}, \bibinfo{person}{Brian Kroth}, \bibinfo{person}{Andreas M\"{u}ller}, \bibinfo{person}{Carlo Curino}, {and} \bibinfo{person}{Shivaram Venkataraman}.} \bibinfo{year}{2022}\natexlab{}.
\newblock \showarticletitle{LlamaTune: sample-efficient DBMS configuration tuning}.
\newblock \bibinfo{journal}{\emph{Proc. VLDB Endow.}} \bibinfo{volume}{15}, \bibinfo{number}{11} (\bibinfo{date}{jul} \bibinfo{year}{2022}), \bibinfo{pages}{2953–2965}.
\newblock
\showISSN{2150-8097}
\urldef\tempurl%
\url{https://doi.org/10.14778/3551793.3551844}
\showDOI{\tempurl}


\bibitem[Kanellis et~al\mbox{.}(2024)]%
        {Nautilus}
\bibfield{author}{\bibinfo{person}{Konstantinos Kanellis}, \bibinfo{person}{Johannes Freischuetz}, {and} \bibinfo{person}{Shivaram Venkataraman}.} \bibinfo{year}{2024}\natexlab{}.
\newblock \showarticletitle{Nautilus: A Benchmarking Platform for DBMS Knob Tuning}. In \bibinfo{booktitle}{\emph{Proceedings of the Eighth Workshop on Data Management for End-to-End Machine Learning}}. \bibinfo{pages}{72--76}.
\newblock


\bibitem[Kavalanekar et~al\mbox{.}(2008)]%
        {WindowsServerDiskVariance}
\bibfield{author}{\bibinfo{person}{Swaroop Kavalanekar}, \bibinfo{person}{Bruce Worthington}, \bibinfo{person}{Qi Zhang}, {and} \bibinfo{person}{Vishal Sharda}.} \bibinfo{year}{2008}\natexlab{}.
\newblock \showarticletitle{Characterization of storage workload traces from production Windows Servers}. In \bibinfo{booktitle}{\emph{2008 IEEE International Symposium on Workload Characterization}}. \bibinfo{pages}{119--128}.
\newblock
\urldef\tempurl%
\url{https://doi.org/10.1109/IISWC.2008.4636097}
\showDOI{\tempurl}


\bibitem[King(2023)]%
        {Stressng}
\bibfield{author}{\bibinfo{person}{Colin~Ian King}.} \bibinfo{year}{2023}\natexlab{}.
\newblock \bibinfo{title}{stress-ng}.
\newblock
\newblock
\newblock
\shownote{\url{https://wiki.ubuntu.com/Kernel/Reference/stress-ng}}.


\bibitem[Konno et~al\mbox{.}(2002)]%
        {MeanSemiVariance2}
\bibfield{author}{\bibinfo{person}{Hiroshi Konno}, \bibinfo{person}{Hayato Waki}, {and} \bibinfo{person}{Atsushi Yuuki}.} \bibinfo{year}{2002}\natexlab{}.
\newblock \showarticletitle{Portfolio optimization under lower partial risk measures}.
\newblock \bibinfo{journal}{\emph{Asia-Pacific Financial Markets}}  \bibinfo{volume}{9} (\bibinfo{year}{2002}), \bibinfo{pages}{127--140}.
\newblock
\urldef\tempurl%
\url{https://doi.org/10.1023/A:1022238119491}
\showURL{%
\tempurl}


\bibitem[Konno and Yamazaki(1991)]%
        {MeanAbsoluteDeviation}
\bibfield{author}{\bibinfo{person}{Hiroshi Konno} {and} \bibinfo{person}{Hiroaki Yamazaki}.} \bibinfo{year}{1991}\natexlab{}.
\newblock \showarticletitle{Mean-absolute deviation portfolio optimization model and its applications to Tokyo stock market}.
\newblock \bibinfo{journal}{\emph{Management science}} \bibinfo{volume}{37}, \bibinfo{number}{5} (\bibinfo{year}{1991}), \bibinfo{pages}{519--531}.
\newblock
\urldef\tempurl%
\url{https://doi.org/10.1287/mnsc.37.5.519}
\showURL{%
\tempurl}


\bibitem[Kopytov(2020)]%
        {sysbench}
\bibfield{author}{\bibinfo{person}{Alexey Kopytov}.} \bibinfo{year}{2020}\natexlab{}.
\newblock \bibinfo{title}{sysbench}.
\newblock
\newblock
\urldef\tempurl%
\url{https://github.com/akopytov/sysbench}
\showURL{%
\tempurl}


\bibitem[Kroth et~al\mbox{.}(2024)]%
        {krothmlos}
\bibfield{author}{\bibinfo{person}{Brian Kroth}, \bibinfo{person}{Sergiy Matusevych}, \bibinfo{person}{Rana Alotaibi}, \bibinfo{person}{Yiwen Zhu}, \bibinfo{person}{Anja Gruenheid}, {and} \bibinfo{person}{Yuanyuan Tian}.} \bibinfo{year}{2024}\natexlab{}.
\newblock \showarticletitle{MLOS in Action: Bridging the Gap Between Experimentation and Auto-Tuning in the Cloud}.
\newblock \bibinfo{journal}{\emph{Proc. VLDB Endow.}} \bibinfo{volume}{17}, \bibinfo{number}{12} (\bibinfo{year}{2024}).
\newblock
\showISSN{2150-8097}
\urldef\tempurl%
\url{https://doi.org/doi:10.14778/3685800.3685852}
\showDOI{\tempurl}


\bibitem[Laaber et~al\mbox{.}(2019)]%
        {howbadistesting}
\bibfield{author}{\bibinfo{person}{Christoph Laaber}, \bibinfo{person}{Joel Scheuner}, {and} \bibinfo{person}{Philipp Leitner}.} \bibinfo{year}{2019}\natexlab{}.
\newblock \showarticletitle{Software microbenchmarking in the cloud. How bad is it really?}
\newblock \bibinfo{journal}{\emph{Empirical Software Engineering}} \bibinfo{volume}{24}, \bibinfo{number}{4} (\bibinfo{year}{2019}), \bibinfo{pages}{2469--2508}.
\newblock
\urldef\tempurl%
\url{https://doi.org/10.1007/s10664-019-09681-1}
\showURL{%
\tempurl}


\bibitem[Lao et~al\mbox{.}(2023)]%
        {gpttuner}
\bibfield{author}{\bibinfo{person}{Jiale Lao}, \bibinfo{person}{Yibo Wang}, \bibinfo{person}{Yufei Li}, \bibinfo{person}{Jianping Wang}, \bibinfo{person}{Yunjia Zhang}, \bibinfo{person}{Zhiyuan Cheng}, \bibinfo{person}{Wanghu Chen}, \bibinfo{person}{Mingjie Tang}, {and} \bibinfo{person}{Jianguo Wang}.} \bibinfo{year}{2023}\natexlab{}.
\newblock \bibinfo{title}{GPTuner: A Manual-Reading Database Tuning System via GPT-Guided Bayesian Optimization}.
\newblock
\newblock
\showeprint[arxiv]{2311.03157}~[cs.DB]
\urldef\tempurl%
\url{https://doi.org/10.14778/3659437.3659449}
\showURL{%
\tempurl}


\bibitem[Leff et~al\mbox{.}(2003)]%
        {define_sla}
\bibfield{author}{\bibinfo{person}{A. Leff}, \bibinfo{person}{J.T. Rayfield}, {and} \bibinfo{person}{D.M. Dias}.} \bibinfo{year}{2003}\natexlab{}.
\newblock \showarticletitle{Service-level agreements and commercial grids}.
\newblock \bibinfo{journal}{\emph{IEEE Internet Computing}} \bibinfo{volume}{7}, \bibinfo{number}{4} (\bibinfo{year}{2003}), \bibinfo{pages}{44--50}.
\newblock
\urldef\tempurl%
\url{https://doi.org/10.1109/MIC.2003.1215659}
\showDOI{\tempurl}


\bibitem[Leis et~al\mbox{.}(2015)]%
        {job_benchmark}
\bibfield{author}{\bibinfo{person}{Viktor Leis}, \bibinfo{person}{Andrey Gubichev}, \bibinfo{person}{Atanas Mirchev}, \bibinfo{person}{Peter Boncz}, \bibinfo{person}{Alfons Kemper}, {and} \bibinfo{person}{Thomas Neumann}.} \bibinfo{year}{2015}\natexlab{}.
\newblock \showarticletitle{How good are query optimizers, really?}
\newblock \bibinfo{journal}{\emph{Proc. VLDB Endow.}} \bibinfo{volume}{9}, \bibinfo{number}{3} (\bibinfo{date}{Nov.} \bibinfo{year}{2015}), \bibinfo{pages}{204–215}.
\newblock
\showISSN{2150-8097}
\urldef\tempurl%
\url{https://doi.org/10.14778/2850583.2850594}
\showDOI{\tempurl}


\bibitem[Leitner and Cito(2016)]%
        {IaaSVariance}
\bibfield{author}{\bibinfo{person}{Philipp Leitner} {and} \bibinfo{person}{J\"{u}rgen Cito}.} \bibinfo{year}{2016}\natexlab{}.
\newblock \showarticletitle{Patterns in the Chaos—A Study of Performance Variation and Predictability in Public IaaS Clouds}.
\newblock \bibinfo{journal}{\emph{ACM Trans. Internet Technol.}} \bibinfo{volume}{16}, \bibinfo{number}{3}, Article \bibinfo{articleno}{15} (\bibinfo{date}{apr} \bibinfo{year}{2016}), \bibinfo{numpages}{23}~pages.
\newblock
\showISSN{1533-5399}
\urldef\tempurl%
\url{https://doi.org/10.1145/2885497}
\showDOI{\tempurl}


\bibitem[Letham et~al\mbox{.}(2019)]%
        {ConstrainedBO}
\bibfield{author}{\bibinfo{person}{Benjamin Letham}, \bibinfo{person}{Brian Karrer}, \bibinfo{person}{Guilherme Ottoni}, {and} \bibinfo{person}{Eytan Bakshy}.} \bibinfo{year}{2019}\natexlab{}.
\newblock \showarticletitle{{Constrained Bayesian Optimization with Noisy Experiments}}.
\newblock \bibinfo{journal}{\emph{Bayesian Analysis}} \bibinfo{volume}{14}, \bibinfo{number}{2} (\bibinfo{year}{2019}), \bibinfo{pages}{495 -- 519}.
\newblock
\urldef\tempurl%
\url{https://doi.org/10.1214/18-BA1110}
\showDOI{\tempurl}
\newblock
\shownote{\url{https://doi.org/10.1214/18-BA1110}}.


\bibitem[Leutenegger and Dias(1993)]%
        {tpcc}
\bibfield{author}{\bibinfo{person}{Scott~T Leutenegger} {and} \bibinfo{person}{Daniel Dias}.} \bibinfo{year}{1993}\natexlab{}.
\newblock \showarticletitle{A modeling study of the TPC-C benchmark}.
\newblock \bibinfo{journal}{\emph{ACM Sigmod Record}} \bibinfo{volume}{22}, \bibinfo{number}{2} (\bibinfo{year}{1993}), \bibinfo{pages}{22--31}.
\newblock
\urldef\tempurl%
\url{http://doi.org/10.1145/170036.170042}
\showURL{%
\tempurl}


\bibitem[Li et~al\mbox{.}(2019)]%
        {QTune}
\bibfield{author}{\bibinfo{person}{Guoliang Li}, \bibinfo{person}{Xuanhe Zhou}, \bibinfo{person}{Shifu Li}, {and} \bibinfo{person}{Bo Gao}.} \bibinfo{year}{2019}\natexlab{}.
\newblock \showarticletitle{QTune: A Query-Aware Database Tuning System with Deep Reinforcement Learning}.
\newblock \bibinfo{journal}{\emph{Proc. VLDB Endow.}} \bibinfo{volume}{12}, \bibinfo{number}{12} (\bibinfo{date}{aug} \bibinfo{year}{2019}), \bibinfo{pages}{2118–2130}.
\newblock
\showISSN{2150-8097}
\urldef\tempurl%
\url{https://doi.org/10.14778/3352063.3352129}
\showDOI{\tempurl}


\bibitem[Lindauer et~al\mbox{.}(2022)]%
        {SMAC3}
\bibfield{author}{\bibinfo{person}{Marius Lindauer}, \bibinfo{person}{Katharina Eggensperger}, \bibinfo{person}{Matthias Feurer}, \bibinfo{person}{André Biedenkapp}, \bibinfo{person}{Difan Deng}, \bibinfo{person}{Carolin Benjamins}, \bibinfo{person}{Tim Ruhkopf}, \bibinfo{person}{René Sass}, {and} \bibinfo{person}{Frank Hutter}.} \bibinfo{year}{2022}\natexlab{}.
\newblock \showarticletitle{SMAC3: A Versatile Bayesian Optimization Package for Hyperparameter Optimization}.
\newblock \bibinfo{journal}{\emph{Journal of Machine Learning Research}} \bibinfo{volume}{23}, \bibinfo{number}{54} (\bibinfo{year}{2022}), \bibinfo{pages}{1--9}.
\newblock
\urldef\tempurl%
\url{http://jmlr.org/papers/v23/21-0888.html}
\showURL{%
\tempurl}
\newblock
\shownote{\url{https://jmlr.org/papers/v23/21-0888.html}}.


\bibitem[Lloyd et~al\mbox{.}(2018)]%
        {MicroserviceVarianceFromServerless}
\bibfield{author}{\bibinfo{person}{Wes Lloyd}, \bibinfo{person}{Shruti Ramesh}, \bibinfo{person}{Swetha Chinthalapati}, \bibinfo{person}{Lan Ly}, {and} \bibinfo{person}{Shrideep Pallickara}.} \bibinfo{year}{2018}\natexlab{}.
\newblock \showarticletitle{Serverless Computing: An Investigation of Factors Influencing Microservice Performance}. In \bibinfo{booktitle}{\emph{2018 IEEE International Conference on Cloud Engineering (IC2E)}}. \bibinfo{publisher}{IEEE}, \bibinfo{address}{Orlando, FL, USA}, \bibinfo{pages}{159--169}.
\newblock
\urldef\tempurl%
\url{https://doi.org/10.1109/IC2E.2018.00039}
\showDOI{\tempurl}


\bibitem[Lorido-Botran et~al\mbox{.}(2017)]%
        {NoisyNeighborDetection1}
\bibfield{author}{\bibinfo{person}{Tania Lorido-Botran}, \bibinfo{person}{Sergio Huerta}, \bibinfo{person}{Luis Tomás}, \bibinfo{person}{Johan Tordsson}, {and} \bibinfo{person}{Borja Sanz}.} \bibinfo{year}{2017}\natexlab{}.
\newblock \showarticletitle{An unsupervised approach to online noisy-neighbor detection in cloud data centers}.
\newblock \bibinfo{journal}{\emph{Expert Systems with Applications}}  \bibinfo{volume}{89} (\bibinfo{year}{2017}), \bibinfo{pages}{188--204}.
\newblock
\showISSN{0957-4174}
\urldef\tempurl%
\url{https://doi.org/10.1016/j.eswa.2017.07.038}
\showDOI{\tempurl}


\bibitem[Mao et~al\mbox{.}(2019)]%
        {clustertuning}
\bibfield{author}{\bibinfo{person}{Hongzi Mao}, \bibinfo{person}{Malte Schwarzkopf}, \bibinfo{person}{Shaileshh~Bojja Venkatakrishnan}, \bibinfo{person}{Zili Meng}, {and} \bibinfo{person}{Mohammad Alizadeh}.} \bibinfo{year}{2019}\natexlab{}.
\newblock \showarticletitle{Learning scheduling algorithms for data processing clusters}. In \bibinfo{booktitle}{\emph{Proceedings of the ACM Special Interest Group on Data Communication}} (Beijing, China) \emph{(\bibinfo{series}{SIGCOMM '19})}. \bibinfo{publisher}{Association for Computing Machinery}, \bibinfo{address}{New York, NY, USA}, \bibinfo{pages}{270–288}.
\newblock
\showISBNx{9781450359566}
\urldef\tempurl%
\url{https://doi.org/10.1145/3341302.3342080}
\showDOI{\tempurl}


\bibitem[Maricq et~al\mbox{.}(2018)]%
        {TamingPerformanceVariability}
\bibfield{author}{\bibinfo{person}{Aleksander Maricq}, \bibinfo{person}{Dmitry Duplyakin}, \bibinfo{person}{Ivo Jimenez}, \bibinfo{person}{Carlos Maltzahn}, \bibinfo{person}{Ryan Stutsman}, {and} \bibinfo{person}{Robert Ricci}.} \bibinfo{year}{2018}\natexlab{}.
\newblock \showarticletitle{Taming Performance Variability}. In \bibinfo{booktitle}{\emph{Proceedings of the 13th USENIX Conference on Operating Systems Design and Implementation}} (Carlsbad, CA, USA) \emph{(\bibinfo{series}{OSDI'18})}. \bibinfo{publisher}{USENIX Association}, \bibinfo{address}{USA}, \bibinfo{pages}{409–425}.
\newblock
\showISBNx{9781931971478}
\urldef\tempurl%
\url{https://dl.acm.org/doi/10.5555/3291168.3291198}
\showURL{%
\tempurl}


\bibitem[Markowitz(1952)]%
        {MarkowitzMeanVariance}
\bibfield{author}{\bibinfo{person}{Harry Markowitz}.} \bibinfo{year}{1952}\natexlab{}.
\newblock \showarticletitle{Portfolio Selection}.
\newblock \bibinfo{journal}{\emph{The Journal of Finance}} \bibinfo{volume}{7}, \bibinfo{number}{1} (\bibinfo{year}{1952}), \bibinfo{pages}{77--91}.
\newblock
\showISSN{00221082, 15406261}
\urldef\tempurl%
\url{http://www.jstor.org/stable/2975974}
\showURL{%
\tempurl}


\bibitem[Oracle(2024)]%
        {OracleCloudOfferings}
Oracle \bibinfo{year}{2024}\natexlab{}.
\newblock \bibinfo{title}{Compute Shapes}.
\newblock
\newblock
\urldef\tempurl%
\url{https://docs.oracle.com/en-us/iaas/Content/Compute/References/computeshapes.htm#vm-standard}
\showURL{%
\tempurl}


\bibitem[Park et~al\mbox{.}(2019)]%
        {MemoryBandwidthParitioning}
\bibfield{author}{\bibinfo{person}{Jinsu Park}, \bibinfo{person}{Seongbeom Park}, {and} \bibinfo{person}{Woongki Baek}.} \bibinfo{year}{2019}\natexlab{}.
\newblock \showarticletitle{CoPart: Coordinated Partitioning of Last-Level Cache and Memory Bandwidth for Fairness-Aware Workload Consolidation on Commodity Servers}. In \bibinfo{booktitle}{\emph{Proceedings of the Fourteenth EuroSys Conference 2019}} (Dresden, Germany) \emph{(\bibinfo{series}{EuroSys '19})}. \bibinfo{publisher}{Association for Computing Machinery}, \bibinfo{address}{New York, NY, USA}, Article \bibinfo{articleno}{10}, \bibinfo{numpages}{16}~pages.
\newblock
\showISBNx{9781450362818}
\urldef\tempurl%
\url{https://doi.org/10.1145/3302424.3303963}
\showDOI{\tempurl}


\bibitem[Poess and Floyd(2000)]%
        {tpch}
\bibfield{author}{\bibinfo{person}{Meikel Poess} {and} \bibinfo{person}{Chris Floyd}.} \bibinfo{year}{2000}\natexlab{}.
\newblock \showarticletitle{New TPC benchmarks for decision support and web commerce}.
\newblock \bibinfo{journal}{\emph{ACM Sigmod Record}} \bibinfo{volume}{29}, \bibinfo{number}{4} (\bibinfo{year}{2000}), \bibinfo{pages}{64--71}.
\newblock
\urldef\tempurl%
\url{http://doi.org/10.1145/369275.369291}
\showURL{%
\tempurl}


\bibitem[Prout et~al\mbox{.}(2018)]%
        {prout2018measuring}
\bibfield{author}{\bibinfo{person}{Andrew Prout}, \bibinfo{person}{William Arcand}, \bibinfo{person}{David Bestor}, \bibinfo{person}{Bill Bergeron}, \bibinfo{person}{Chansup Byun}, \bibinfo{person}{Vijay Gadepally}, \bibinfo{person}{Michael Houle}, \bibinfo{person}{Matthew Hubbell}, \bibinfo{person}{Michael Jones}, \bibinfo{person}{Anna Klein}, {et~al\mbox{.}}} \bibinfo{year}{2018}\natexlab{}.
\newblock \showarticletitle{Measuring the impact of spectre and meltdown}. In \bibinfo{booktitle}{\emph{2018 IEEE High Performance extreme Computing Conference (HPEC)}}. IEEE, \bibinfo{pages}{1--5}.
\newblock


\bibitem[Pu et~al\mbox{.}(2013)]%
        {CloudIOVariance}
\bibfield{author}{\bibinfo{person}{Xing Pu}, \bibinfo{person}{Ling Liu}, \bibinfo{person}{Yiduo Mei}, \bibinfo{person}{Sankaran Sivathanu}, \bibinfo{person}{Younggyun Koh}, \bibinfo{person}{Calton Pu}, {and} \bibinfo{person}{Yuanda Cao}.} \bibinfo{year}{2013}\natexlab{}.
\newblock \showarticletitle{Who Is Your Neighbor: Net I/O Performance Interference in Virtualized Clouds}.
\newblock \bibinfo{journal}{\emph{IEEE Transactions on Services Computing}} \bibinfo{volume}{6}, \bibinfo{number}{3} (\bibinfo{year}{2013}), \bibinfo{pages}{314--329}.
\newblock
\urldef\tempurl%
\url{https://doi.org/10.1109/TSC.2012.2}
\showDOI{\tempurl}


\bibitem[Qiu et~al\mbox{.}(2020)]%
        {FIRM}
\bibfield{author}{\bibinfo{person}{Haoran Qiu}, \bibinfo{person}{Subho~S. Banerjee}, \bibinfo{person}{Saurabh Jha}, \bibinfo{person}{Zbigniew~T. Kalbarczyk}, {and} \bibinfo{person}{Ravishankar~K. Iyer}.} \bibinfo{year}{2020}\natexlab{}.
\newblock \showarticletitle{{FIRM}: An Intelligent Fine-grained Resource Management Framework for {SLO-Oriented} Microservices}. In \bibinfo{booktitle}{\emph{14th USENIX Symposium on Operating Systems Design and Implementation (OSDI 20)}}. \bibinfo{publisher}{USENIX Association}, \bibinfo{pages}{805--825}.
\newblock
\showISBNx{978-1-939133-19-9}
\urldef\tempurl%
\url{https://www.usenix.org/conference/osdi20/presentation/qiu}
\showURL{%
\tempurl}


\bibitem[Richardson et~al\mbox{.}(2003)]%
        {epinions}
\bibfield{author}{\bibinfo{person}{Matthew Richardson}, \bibinfo{person}{Rakesh Agrawal}, {and} \bibinfo{person}{Pedro Domingos}.} \bibinfo{year}{2003}\natexlab{}.
\newblock \showarticletitle{Trust management for the semantic web}. In \bibinfo{booktitle}{\emph{International semantic Web conference}}. Springer, \bibinfo{pages}{351--368}.
\newblock
\newblock
\shownote{\url{https://doi.org/10.1007/978-3-540-39718-2_23}}.


\bibitem[Rodola(2024)]%
        {psutil}
\bibfield{author}{\bibinfo{person}{Giampaolo Rodola}.} \bibinfo{year}{2024}\natexlab{}.
\newblock \bibinfo{title}{psutil}.
\newblock
\newblock
\urldef\tempurl%
\url{https://github.com/giampaolo/psutil}
\showURL{%
\tempurl}


\bibitem[Roygara et~al\mbox{.}(2024)]%
        {ssdv2}
\bibfield{author}{\bibinfo{person}{Roygara}, \bibinfo{person}{Masha Thomas}, \bibinfo{person}{Becca Goodman}, \bibinfo{person}{Jeff Borsecnik}, \bibinfo{person}{Henry Hagnäs}, \bibinfo{person}{Raman Kumar}, \bibinfo{person}{Sumanth Marigowda}, \bibinfo{person}{Stephen Haas}, \bibinfo{person}{Micah McKittrick}, \bibinfo{person}{David Coulter}, {and} \bibinfo{person}{et al.}} \bibinfo{year}{2024}\natexlab{}.
\newblock \bibinfo{title}{Azure managed disk types}.
\newblock
\newblock
\urldef\tempurl%
\url{https://learn.microsoft.com/en-us/azure/virtual-machines/disks-types#premium-ssd-v2}
\showURL{%
\tempurl}


\bibitem[Samuelson(1970)]%
        {MeanVarianceSkewModel}
\bibfield{author}{\bibinfo{person}{Paul~A. Samuelson}.} \bibinfo{year}{1970}\natexlab{}.
\newblock \showarticletitle{{The Fundamental Approximation Theorem of Portfolio Analysis in terms of Means, Variances and Higher Moments1}}.
\newblock \bibinfo{journal}{\emph{The Review of Economic Studies}} \bibinfo{volume}{37}, \bibinfo{number}{4} (\bibinfo{date}{10} \bibinfo{year}{1970}), \bibinfo{pages}{537--542}.
\newblock
\showISSN{0034-6527}
\urldef\tempurl%
\url{https://doi.org/10.2307/2296483}
\showDOI{\tempurl}
\showeprint{https://academic.oup.com/restud/article-pdf/37/4/537/4353167/37-4-537.pdf}


\bibitem[Sanfilippo(2022)]%
        {redisbenchmark}
\bibfield{author}{\bibinfo{person}{Salvatore Sanfilippo}.} \bibinfo{year}{2022}\natexlab{}.
\newblock \bibinfo{title}{Redis Benchmark}.
\newblock
\newblock
\urldef\tempurl%
\url{https://redis.io/docs/management/optimization/benchmarks/}
\showURL{%
\tempurl}


\bibitem[Schad et~al\mbox{.}(2010)]%
        {EC2PerfVariance}
\bibfield{author}{\bibinfo{person}{J\"{o}rg Schad}, \bibinfo{person}{Jens Dittrich}, {and} \bibinfo{person}{Jorge-Arnulfo Quian\'{e}-Ruiz}.} \bibinfo{year}{2010}\natexlab{}.
\newblock \showarticletitle{Runtime Measurements in the Cloud: Observing, Analyzing, and Reducing Variance}.
\newblock \bibinfo{journal}{\emph{Proc. VLDB Endow.}} \bibinfo{volume}{3}, \bibinfo{number}{1–2} (\bibinfo{date}{sep} \bibinfo{year}{2010}), \bibinfo{pages}{460–471}.
\newblock
\showISSN{2150-8097}
\urldef\tempurl%
\url{https://doi.org/10.14778/1920841.1920902}
\showDOI{\tempurl}


\bibitem[Scheuner and Leitner(2018)]%
        {CloudBenchmarkSuite}
\bibfield{author}{\bibinfo{person}{Joel Scheuner} {and} \bibinfo{person}{Philipp Leitner}.} \bibinfo{year}{2018}\natexlab{}.
\newblock \showarticletitle{A Cloud Benchmark Suite Combining Micro and Applications Benchmarks}. In \bibinfo{booktitle}{\emph{Companion of the 2018 ACM/SPEC International Conference on Performance Engineering}} (Berlin, Germany) \emph{(\bibinfo{series}{ICPE '18})}. \bibinfo{publisher}{Association for Computing Machinery}, \bibinfo{address}{New York, NY, USA}, \bibinfo{pages}{161–166}.
\newblock
\showISBNx{9781450356299}
\urldef\tempurl%
\url{https://doi.org/10.1145/3185768.3186286}
\showDOI{\tempurl}


\bibitem[Schiefer and Valentin(1999)]%
        {db2_tuning_no_ml}
\bibfield{author}{\bibinfo{person}{K.~Bernhard Schiefer} {and} \bibinfo{person}{Gary Valentin}.} \bibinfo{year}{1999}\natexlab{}.
\newblock \showarticletitle{DB2 universal database performance tuning}.
\newblock \bibinfo{journal}{\emph{IEEE Data Eng. Bull.}} \bibinfo{volume}{22}, \bibinfo{number}{2} (\bibinfo{year}{1999}), \bibinfo{pages}{12--19}.
\newblock


\bibitem[Segal(2004)]%
        {RandomForestBenchmark}
\bibfield{author}{\bibinfo{person}{Mark~R Segal}.} \bibinfo{year}{2004}\natexlab{}.
\newblock \showarticletitle{Machine learning benchmarks and random forest regression}.
\newblock \bibinfo{journal}{\emph{CSF: Center for Bioinformatics and Molecular Biostatistics}} (\bibinfo{year}{2004}).
\newblock


\bibitem[Sinha et~al\mbox{.}(2022)]%
        {GPUVariance}
\bibfield{author}{\bibinfo{person}{Prasoon Sinha}, \bibinfo{person}{Akhil Guliani}, \bibinfo{person}{Rutwik Jain}, \bibinfo{person}{Brandon Tran}, \bibinfo{person}{Matthew~D. Sinclair}, {and} \bibinfo{person}{Shivaram Venkataraman}.} \bibinfo{year}{2022}\natexlab{}.
\newblock \showarticletitle{Not All GPUs Are Created Equal: Characterizing Variability in Large-Scale, Accelerator-Rich Systems}. In \bibinfo{booktitle}{\emph{Proceedings of the International Conference on High Performance Computing, Networking, Storage and Analysis}} (Dallas, Texas) \emph{(\bibinfo{series}{SC '22})}. \bibinfo{publisher}{IEEE Press}, \bibinfo{address}{Dallas, Texas}, Article \bibinfo{articleno}{65}, \bibinfo{numpages}{15}~pages.
\newblock
\showISBNx{9784665454445}
\urldef\tempurl%
\url{https://dl.acm.org/doi/abs/10.5555/3571885.3571971}
\showURL{%
\tempurl}


\bibitem[Skinner and Kramer(2005)]%
        {HPCVariability}
\bibfield{author}{\bibinfo{person}{D. Skinner} {and} \bibinfo{person}{W. Kramer}.} \bibinfo{year}{2005}\natexlab{}.
\newblock \showarticletitle{Understanding the causes of performance variability in HPC workloads}. In \bibinfo{booktitle}{\emph{IEEE International. 2005 Proceedings of the IEEE Workload Characterization Symposium, 2005.}} \bibinfo{publisher}{IEEE Press}, \bibinfo{address}{Austin, TX}, \bibinfo{pages}{137--149}.
\newblock
\urldef\tempurl%
\url{https://doi.org/10.1109/IISWC.2005.1526010}
\showDOI{\tempurl}


\bibitem[Stonebraker and Rowe(1986)]%
        {postgres}
\bibfield{author}{\bibinfo{person}{Michael Stonebraker} {and} \bibinfo{person}{Lawrence~A. Rowe}.} \bibinfo{year}{1986}\natexlab{}.
\newblock \showarticletitle{The design of POSTGRES}.
\newblock \bibinfo{journal}{\emph{SIGMOD Rec.}} \bibinfo{volume}{15}, \bibinfo{number}{2} (\bibinfo{date}{June} \bibinfo{year}{1986}), \bibinfo{pages}{340–355}.
\newblock
\showISSN{0163-5808}
\urldef\tempurl%
\url{https://doi.org/10.1145/16856.16888}
\showDOI{\tempurl}


\bibitem[Thompson(1933)]%
        {tohmpson_sampling}
\bibfield{author}{\bibinfo{person}{William~R. Thompson}.} \bibinfo{year}{1933}\natexlab{}.
\newblock \showarticletitle{On the Likelihood that One Unknown Probability Exceeds Another in View of the Evidence of Two Samples}.
\newblock \bibinfo{journal}{\emph{Biometrika}} \bibinfo{volume}{25}, \bibinfo{number}{3/4} (\bibinfo{year}{1933}), \bibinfo{pages}{285--294}.
\newblock
\showISSN{00063444}
\urldef\tempurl%
\url{http://www.jstor.org/stable/2332286}
\showURL{%
\tempurl}


\bibitem[Tiwari et~al\mbox{.}(2009)]%
        {compiler_tuning}
\bibfield{author}{\bibinfo{person}{Ananta Tiwari}, \bibinfo{person}{Chun Chen}, \bibinfo{person}{Jacqueline Chame}, \bibinfo{person}{Mary Hall}, {and} \bibinfo{person}{Jeffrey~K. Hollingsworth}.} \bibinfo{year}{2009}\natexlab{}.
\newblock \showarticletitle{A scalable auto-tuning framework for compiler optimization}. In \bibinfo{booktitle}{\emph{2009 IEEE International Symposium on Parallel \& Distributed Processing}}. \bibinfo{pages}{1--12}.
\newblock
\urldef\tempurl%
\url{https://doi.org/10.1109/IPDPS.2009.5161054}
\showDOI{\tempurl}


\bibitem[Track(2020)]%
        {mssales}
\bibfield{author}{\bibinfo{person}{Inside Track}.} \bibinfo{year}{2020}\natexlab{}.
\newblock
\newblock
\urldef\tempurl%
\url{https://www.microsoft.com/insidetrack/blog/microsoft-reinvents-sales-processing-and-financial-reporting-with-azure/}
\showURL{%
\tempurl}


\bibitem[Trummer(2022)]%
        {DBBert}
\bibfield{author}{\bibinfo{person}{Immanuel Trummer}.} \bibinfo{year}{2022}\natexlab{}.
\newblock \showarticletitle{DB-BERT: A Database Tuning Tool that "Reads the Manual"}. In \bibinfo{booktitle}{\emph{Proceedings of the 2022 International Conference on Management of Data}} (Philadelphia, PA, USA) \emph{(\bibinfo{series}{SIGMOD '22})}. \bibinfo{publisher}{Association for Computing Machinery}, \bibinfo{address}{New York, NY, USA}, \bibinfo{pages}{190–203}.
\newblock
\showISBNx{9781450392495}
\urldef\tempurl%
\url{https://doi.org/10.1145/3514221.3517843}
\showDOI{\tempurl}


\bibitem[Uhlig et~al\mbox{.}(2005)]%
        {IntelVtX}
\bibfield{author}{\bibinfo{person}{R. Uhlig}, \bibinfo{person}{G. Neiger}, \bibinfo{person}{D. Rodgers}, \bibinfo{person}{A.L. Santoni}, \bibinfo{person}{F.C.M. Martins}, \bibinfo{person}{A.V. Anderson}, \bibinfo{person}{S.M. Bennett}, \bibinfo{person}{A. Kagi}, \bibinfo{person}{F.H. Leung}, {and} \bibinfo{person}{L. Smith}.} \bibinfo{year}{2005}\natexlab{}.
\newblock \showarticletitle{Intel virtualization technology}.
\newblock \bibinfo{journal}{\emph{Computer}} \bibinfo{volume}{38}, \bibinfo{number}{5} (\bibinfo{year}{2005}), \bibinfo{pages}{48--56}.
\newblock
\urldef\tempurl%
\url{https://doi.org/10.1109/MC.2005.163}
\showDOI{\tempurl}


\bibitem[Uta et~al\mbox{.}(2020)]%
        {bigdatareproducable}
\bibfield{author}{\bibinfo{person}{Alexandru Uta}, \bibinfo{person}{Alexandru Custura}, \bibinfo{person}{Dmitry Duplyakin}, \bibinfo{person}{Ivo Jimenez}, \bibinfo{person}{Jan Rellermeyer}, \bibinfo{person}{Carlos Maltzahn}, \bibinfo{person}{Robert Ricci}, {and} \bibinfo{person}{Alexandru Iosup}.} \bibinfo{year}{2020}\natexlab{}.
\newblock \showarticletitle{Is Big Data Performance Reproducible in Modern Cloud Networks?}. In \bibinfo{booktitle}{\emph{17th USENIX Symposium on Networked Systems Design and Implementation (NSDI 20)}}. \bibinfo{publisher}{USENIX Association}, \bibinfo{address}{Santa Clara, CA}, \bibinfo{pages}{513--527}.
\newblock
\showISBNx{978-1-939133-13-7}
\urldef\tempurl%
\url{https://www.usenix.org/conference/nsdi20/presentation/uta}
\showURL{%
\tempurl}


\bibitem[Van~Aken et~al\mbox{.}(2021)]%
        {OtterTuneRealWorld}
\bibfield{author}{\bibinfo{person}{Dana Van~Aken}, \bibinfo{person}{Dongsheng Yang}, \bibinfo{person}{Sebastien Brillard}, \bibinfo{person}{Ari Fiorino}, \bibinfo{person}{Bohan Zhang}, \bibinfo{person}{Christian Bilien}, {and} \bibinfo{person}{Andrew Pavlo}.} \bibinfo{year}{2021}\natexlab{}.
\newblock \showarticletitle{An Inquiry into Machine Learning-Based Automatic Configuration Tuning Services on Real-World Database Management Systems}.
\newblock \bibinfo{journal}{\emph{Proc. VLDB Endow.}} \bibinfo{volume}{14}, \bibinfo{number}{7} (\bibinfo{date}{apr} \bibinfo{year}{2021}), \bibinfo{pages}{1241–1253}.
\newblock
\showISSN{2150-8097}
\urldef\tempurl%
\url{https://doi.org/10.14778/3450980.3450992}
\showDOI{\tempurl}


\bibitem[Veitch et~al\mbox{.}(2017)]%
        {IntelNFV}
\bibfield{author}{\bibinfo{person}{Paul Veitch}, \bibinfo{person}{Edel Curley}, {and} \bibinfo{person}{Tomasz Kantecki}.} \bibinfo{year}{2017}\natexlab{}.
\newblock \showarticletitle{Performance evaluation of cache allocation technology for NFV noisy neighbor mitigation}. In \bibinfo{booktitle}{\emph{2017 IEEE Conference on Network Softwarization (NetSoft)}}. \bibinfo{pages}{1--5}.
\newblock
\urldef\tempurl%
\url{https://doi.org/10.1109/NETSOFT.2017.8004214}
\showDOI{\tempurl}


\bibitem[Verma et~al\mbox{.}(2024)]%
        {B_Series}
\bibfield{author}{\bibinfo{person}{Rishab Verma}, \bibinfo{person}{Jeff Borsecnik}, \bibinfo{person}{Willie Williams}, \bibinfo{person}{Aarthi Vijayaraghavan}, \bibinfo{person}{Micah Mckittrick}, \bibinfo{person}{Styli Tsinaroglou}, \bibinfo{person}{Justin Chizer}, \bibinfo{person}{Stephen Haas}, \bibinfo{person}{Alec}, \bibinfo{person}{Zach Olinske}, {and} \bibinfo{person}{et al.}} \bibinfo{year}{2024}\natexlab{}.
\newblock \bibinfo{title}{B-series burstable virtual machine sizes}.
\newblock
\newblock
\urldef\tempurl%
\url{https://learn.microsoft.com/en-us/azure/virtual-machines/sizes-b-series-burstable}
\showURL{%
\tempurl}


\bibitem[Viswanathan et~al\mbox{.}(2021)]%
        {IntelMemoryLatencyChecker}
\bibfield{author}{\bibinfo{person}{Vish Viswanathan}, \bibinfo{person}{Karthik Kumar}, \bibinfo{person}{Thomas Willhalm}, {and} \bibinfo{person}{Sri Sakthivelu}.} \bibinfo{year}{2021}\natexlab{}.
\newblock \bibinfo{title}{Intel Memory Latency Checker}.
\newblock
\newblock
\urldef\tempurl%
\url{https://www.intel.com/content/www/us/en/developer/articles/tool/intelr-memory-latency-checker.html}
\showURL{%
\tempurl}
\newblock
\shownote{\url{https://www.intel.com/content/www/us/en/developer/articles/tool/intelr-memory-latency-checker.html}}.


\bibitem[Wang et~al\mbox{.}(2010a)]%
        {outlierdetection1}
\bibfield{author}{\bibinfo{person}{Chengwei Wang}, \bibinfo{person}{Vanish Talwar}, \bibinfo{person}{Karsten Schwan}, {and} \bibinfo{person}{Parthasarathy Ranganathan}.} \bibinfo{year}{2010}\natexlab{a}.
\newblock \showarticletitle{Online detection of utility cloud anomalies using metric distributions}. In \bibinfo{booktitle}{\emph{2010 IEEE Network Operations and Management Symposium - NOMS 2010}}. \bibinfo{pages}{96--103}.
\newblock
\urldef\tempurl%
\url{https://doi.org/10.1109/NOMS.2010.5488443}
\showDOI{\tempurl}


\bibitem[Wang et~al\mbox{.}(2010b)]%
        {NoisyNeighborDetetion2}
\bibfield{author}{\bibinfo{person}{Chengwei Wang}, \bibinfo{person}{Vanish Talwar}, \bibinfo{person}{Karsten Schwan}, {and} \bibinfo{person}{Parthasarathy Ranganathan}.} \bibinfo{year}{2010}\natexlab{b}.
\newblock \showarticletitle{Online detection of utility cloud anomalies using metric distributions}. In \bibinfo{booktitle}{\emph{2010 IEEE Network Operations and Management Symposium - NOMS 2010}}. \bibinfo{pages}{96--103}.
\newblock
\urldef\tempurl%
\url{https://doi.org/10.1109/NOMS.2010.5488443}
\showDOI{\tempurl}


\bibitem[Wang et~al\mbox{.}(2016)]%
        {MLScheduler}
\bibfield{author}{\bibinfo{person}{Guolu Wang}, \bibinfo{person}{Jungang Xu}, {and} \bibinfo{person}{Ben He}.} \bibinfo{year}{2016}\natexlab{}.
\newblock \showarticletitle{A Novel Method for Tuning Configuration Parameters of Spark Based on Machine Learning}. In \bibinfo{booktitle}{\emph{2016 IEEE 18th International Conference on High Performance Computing and Communications; IEEE 14th International Conference on Smart City; IEEE 2nd International Conference on Data Science and Systems (HPCC/SmartCity/DSS)}}. \bibinfo{pages}{586--593}.
\newblock
\urldef\tempurl%
\url{https://doi.org/10.1109/HPCC-SmartCity-DSS.2016.0088}
\showDOI{\tempurl}
\newblock
\shownote{\url{http://doi.org/10.1109/HPCC-SmartCity-DSS.2016.0088}}.


\bibitem[Yadwadkar et~al\mbox{.}(2017)]%
        {select_best_vm}
\bibfield{author}{\bibinfo{person}{Neeraja~J. Yadwadkar}, \bibinfo{person}{Bharath Hariharan}, \bibinfo{person}{Joseph~E. Gonzalez}, \bibinfo{person}{Burton Smith}, {and} \bibinfo{person}{Randy~H. Katz}.} \bibinfo{year}{2017}\natexlab{}.
\newblock \showarticletitle{Selecting the best VM across multiple public clouds: a data-driven performance modeling approach}. In \bibinfo{booktitle}{\emph{Proceedings of the 2017 Symposium on Cloud Computing}} (Santa Clara, California) \emph{(\bibinfo{series}{SoCC '17})}. \bibinfo{publisher}{Association for Computing Machinery}, \bibinfo{address}{New York, NY, USA}, \bibinfo{pages}{452–465}.
\newblock
\showISBNx{9781450350280}
\urldef\tempurl%
\url{https://doi.org/10.1145/3127479.3131614}
\showDOI{\tempurl}


\bibitem[Yoo et~al\mbox{.}(2021)]%
        {GrayFailure2}
\bibfield{author}{\bibinfo{person}{Andrew Yoo}, \bibinfo{person}{Yuanli Wang}, \bibinfo{person}{Ritesh Sinha}, \bibinfo{person}{Shuai Mu}, {and} \bibinfo{person}{Tianyin Xu}.} \bibinfo{year}{2021}\natexlab{}.
\newblock \showarticletitle{Fail-Slow Fault Tolerance Needs Programming Support}. In \bibinfo{booktitle}{\emph{Proceedings of the Workshop on Hot Topics in Operating Systems}} (Ann Arbor, Michigan) \emph{(\bibinfo{series}{HotOS '21})}. \bibinfo{publisher}{Association for Computing Machinery}, \bibinfo{address}{New York, NY, USA}, \bibinfo{pages}{228–235}.
\newblock
\showISBNx{9781450384384}
\urldef\tempurl%
\url{https://doi.org/10.1145/3458336.3465299}
\showDOI{\tempurl}


\bibitem[Yuan et~al\mbox{.}(2021)]%
        {IntelDDIO}
\bibfield{author}{\bibinfo{person}{Yifan Yuan}, \bibinfo{person}{Mohammad Alian}, \bibinfo{person}{Yipeng Wang}, \bibinfo{person}{Ren Wang}, \bibinfo{person}{Ilia Kurakin}, \bibinfo{person}{Charlie Tai}, {and} \bibinfo{person}{Nam~Sung Kim}.} \bibinfo{year}{2021}\natexlab{}.
\newblock \showarticletitle{Don’t Forget the I/O When Allocating Your LLC}. In \bibinfo{booktitle}{\emph{2021 ACM/IEEE 48th Annual International Symposium on Computer Architecture (ISCA)}}. \bibinfo{pages}{112--125}.
\newblock
\urldef\tempurl%
\url{https://doi.org/10.1109/ISCA52012.2021.00018}
\showDOI{\tempurl}


\bibitem[Zhang et~al\mbox{.}(2018)]%
        {Ottertune}
\bibfield{author}{\bibinfo{person}{Bohan Zhang}, \bibinfo{person}{Dana Van~Aken}, \bibinfo{person}{Justin Wang}, \bibinfo{person}{Tao Dai}, \bibinfo{person}{Shuli Jiang}, \bibinfo{person}{Jacky Lao}, \bibinfo{person}{Siyuan Sheng}, \bibinfo{person}{Andrew Pavlo}, {and} \bibinfo{person}{Geoffrey~J. Gordon}.} \bibinfo{year}{2018}\natexlab{}.
\newblock \showarticletitle{A Demonstration of the Ottertune Automatic Database Management System Tuning Service}.
\newblock \bibinfo{journal}{\emph{Proc. VLDB Endow.}} \bibinfo{volume}{11}, \bibinfo{number}{12} (\bibinfo{date}{aug} \bibinfo{year}{2018}), \bibinfo{pages}{1910–1913}.
\newblock
\showISSN{2150-8097}
\urldef\tempurl%
\url{https://doi.org/10.14778/3229863.3236222}
\showDOI{\tempurl}


\bibitem[Zhang et~al\mbox{.}(2019)]%
        {CDBTune}
\bibfield{author}{\bibinfo{person}{Ji Zhang}, \bibinfo{person}{Yu Liu}, \bibinfo{person}{Ke Zhou}, \bibinfo{person}{Guoliang Li}, \bibinfo{person}{Zhili Xiao}, \bibinfo{person}{Bin Cheng}, \bibinfo{person}{Jiashu Xing}, \bibinfo{person}{Yangtao Wang}, \bibinfo{person}{Tianheng Cheng}, \bibinfo{person}{Li Liu}, \bibinfo{person}{Minwei Ran}, {and} \bibinfo{person}{Zekang Li}.} \bibinfo{year}{2019}\natexlab{}.
\newblock \showarticletitle{An End-to-End Automatic Cloud Database Tuning System Using Deep Reinforcement Learning}. In \bibinfo{booktitle}{\emph{Proceedings of the 2019 International Conference on Management of Data}} (Amsterdam, Netherlands) \emph{(\bibinfo{series}{SIGMOD '19})}. \bibinfo{publisher}{Association for Computing Machinery}, \bibinfo{address}{New York, NY, USA}, \bibinfo{pages}{415–432}.
\newblock
\showISBNx{9781450356435}
\urldef\tempurl%
\url{https://doi.org/10.1145/3299869.3300085}
\showDOI{\tempurl}


\bibitem[Zhang et~al\mbox{.}(2022a)]%
        {FacilitatingDBMSTuning}
\bibfield{author}{\bibinfo{person}{Xinyi Zhang}, \bibinfo{person}{Zhuo Chang}, \bibinfo{person}{Yang Li}, \bibinfo{person}{Hong Wu}, \bibinfo{person}{Jian Tan}, \bibinfo{person}{Feifei Li}, {and} \bibinfo{person}{Bin Cui}.} \bibinfo{year}{2022}\natexlab{a}.
\newblock \showarticletitle{Facilitating Database Tuning with Hyper-Parameter Optimization: A Comprehensive Experimental Evaluation}.
\newblock \bibinfo{journal}{\emph{Proc. VLDB Endow.}} \bibinfo{volume}{15}, \bibinfo{number}{9} (\bibinfo{date}{jul} \bibinfo{year}{2022}), \bibinfo{pages}{1808–1821}.
\newblock
\showISSN{2150-8097}
\urldef\tempurl%
\url{https://doi.org/10.14778/3538598.3538604}
\showDOI{\tempurl}


\bibitem[Zhang et~al\mbox{.}(2013)]%
        {CPUIsolationGoogle}
\bibfield{author}{\bibinfo{person}{Xiao Zhang}, \bibinfo{person}{Eric Tune}, \bibinfo{person}{Robert Hagmann}, \bibinfo{person}{Rohit Jnagal}, \bibinfo{person}{Vrigo Gokhale}, {and} \bibinfo{person}{John Wilkes}.} \bibinfo{year}{2013}\natexlab{}.
\newblock \showarticletitle{CPI2: CPU performance isolation for shared compute clusters}. In \bibinfo{booktitle}{\emph{Proceedings of the 8th ACM European Conference on Computer Systems}} (Prague, Czech Republic) \emph{(\bibinfo{series}{EuroSys '13})}. \bibinfo{publisher}{Association for Computing Machinery}, \bibinfo{address}{New York, NY, USA}, \bibinfo{pages}{379–391}.
\newblock
\showISBNx{9781450319942}
\urldef\tempurl%
\url{https://doi.org/10.1145/2465351.2465388}
\showDOI{\tempurl}


\bibitem[Zhang et~al\mbox{.}(2022b)]%
        {OnlineTune}
\bibfield{author}{\bibinfo{person}{Xinyi Zhang}, \bibinfo{person}{Hong Wu}, \bibinfo{person}{Yang Li}, \bibinfo{person}{Jian Tan}, \bibinfo{person}{Feifei Li}, {and} \bibinfo{person}{Bin Cui}.} \bibinfo{year}{2022}\natexlab{b}.
\newblock \showarticletitle{Towards Dynamic and Safe Configuration Tuning for Cloud Databases}. In \bibinfo{booktitle}{\emph{Proceedings of the 2022 International Conference on Management of Data}} (Philadelphia, PA, USA) \emph{(\bibinfo{series}{SIGMOD '22})}. \bibinfo{publisher}{Association for Computing Machinery}, \bibinfo{address}{New York, NY, USA}, \bibinfo{pages}{631–645}.
\newblock
\showISBNx{9781450392495}
\urldef\tempurl%
\url{https://doi.org/10.1145/3514221.3526176}
\showDOI{\tempurl}


\bibitem[Zhang et~al\mbox{.}(2024)]%
        {IntelHDIOV}
\bibfield{author}{\bibinfo{person}{Zongpu Zhang}, \bibinfo{person}{Jiangtao Chen}, \bibinfo{person}{Banghao Ying}, \bibinfo{person}{Yahui Cao}, \bibinfo{person}{Lingyu Liu}, \bibinfo{person}{Jian Li}, \bibinfo{person}{Xin Zeng}, \bibinfo{person}{Junyuan Wang}, \bibinfo{person}{Weigang Li}, {and} \bibinfo{person}{Haibing Guan}.} \bibinfo{year}{2024}\natexlab{}.
\newblock \showarticletitle{HD-IOV: SW-HW Co-designed I/O Virtualization with Scalability and Flexibility for Hyper-Density Cloud}. In \bibinfo{booktitle}{\emph{Proceedings of the Nineteenth European Conference on Computer Systems}} (Athens, Greece) \emph{(\bibinfo{series}{EuroSys '24})}. \bibinfo{publisher}{Association for Computing Machinery}, \bibinfo{address}{New York, NY, USA}, \bibinfo{pages}{834–850}.
\newblock
\showISBNx{9798400704376}
\urldef\tempurl%
\url{https://doi.org/10.1145/3627703.3629557}
\showDOI{\tempurl}


\bibitem[Zuck et~al\mbox{.}(2019)]%
        {SSDPerfTransperency}
\bibfield{author}{\bibinfo{person}{Aviad Zuck}, \bibinfo{person}{Philipp G\"{u}hring}, \bibinfo{person}{Tao Zhang}, \bibinfo{person}{Donald~E. Porter}, {and} \bibinfo{person}{Dan Tsafrir}.} \bibinfo{year}{2019}\natexlab{}.
\newblock \showarticletitle{Why and How to Increase SSD Performance Transparency}. In \bibinfo{booktitle}{\emph{Proceedings of the Workshop on Hot Topics in Operating Systems}} (Bertinoro, Italy) \emph{(\bibinfo{series}{HotOS '19})}. \bibinfo{publisher}{Association for Computing Machinery}, \bibinfo{address}{New York, NY, USA}, \bibinfo{pages}{192–200}.
\newblock
\showISBNx{9781450367271}
\urldef\tempurl%
\url{https://doi.org/10.1145/3317550.3321430}
\showDOI{\tempurl}


\end{thebibliography}

\begin{appendices}
%

\section{Artifact Appendix} 
\subsection{Abstract}
In our artifact, we release the source code used for "TUNA: Tuning Unstable and Noisy Cloud Applications", the data from the tuning runs, and the longitudinal data used in the motivation.
The code can be used to run the main experimental results from this paper.
Users interested in using these techniques for their own works can make small modifications to the provided scripts to integrate their systems.

\subsection{Description \& Requirements}

\subsubsection{How to access}
We have released the code for running \SystemName on GitHub~\footnote{\url{https://aka.ms/mlos/tuna-eurosys-artifacts}}.
The data from the tuning runs will also be available here.
Detailed instructions on how to run the experiment can be found on GitHub.
We have also released the longitudinal study datasets at~\footnote{\url{https://aka.ms/mlos/tuna-eurosys-dataset}}.
This includes all of the data required to make the figures for motivation.
These repositories can simply be copied from the respective code bases.
Note, that we use submodules, so we recommend pulling using the instructions listed in the repository.

\subsubsection{Hardware dependencies}
The experiments in this paper were run on Azure and CloudLab. 
We specifically ran D8s\_v5 VMs on Azure, and c220g5 nodes on CloudLab.

\subsubsection{Software dependencies}
We have a relatively long list of dependencies that are described in the GitHub README file.
There are scripts that will automatically install them into a virtual environment on all nodes in the cluster.

\subsubsection{Benchmarks} 
We use a series of benchmarks from Benchbase~\cite{Benchbase} and YCSB~\cite{YCSB}.
These are managed by \texttt{nautilus}, and are automatically installed through the \texttt{docker} containers.

\subsection{Setup}
The environment for TUNA requires two different types of nodes: an Orchestrator and a set of Workers.
These two different types of nodes have slightly different setup instructions, found below.
All scripts are found in \texttt{src/processing}.
One script that may be useful is 

\texttt{add\_hosts.sh <hosts> <port>}

as a way to add all of the hosts listed in the specified host file to the trusted hosts list.
This is required for \texttt{pssh} which we use to deploy our scripts.
One point of note is that you cannot have your username in the host file when using this script.
These scripts should work on any platform, however, we have only tested this on Azure, and CloudLab.
We provide experiment files for c220g5 nodes on CloudLab, and D8s\_v5 nodes in Azure.
These instance types will become important later, however users can provide their own files in \texttt{/src/spaces}.
If you are trying to replicate the work found in our paper, we recommend using 10 worker nodes and 1 orchestrator node, where the 10 worker nodes are the first 10 nodes that were created.
This will allow you to use our provided hosts files (\texttt{hosts.azure} or \texttt{hosts.cloudlab}).
Alternatively, a custom host file can be used.

\subsubsection{Workers}
To install and copy our files, there are two commands we will need to run.

\texttt{./worker\_setup\_remote.sh <hosts>}

The first command will install all of the dependencies, as well as set up the environment.

\texttt{./worker\_deployment.sh <hosts> <node\_type>}

The second command will start all of the required processes.
Note that the second command will say some of the commands fail.
This is expected, as they simply ensure that any previous instances of stopped and deleted before beginning the initialization process.

At some point during running this command, there will be a required interaction to specify that the docker image that is building in the background has been completed.
There are two options here.
First, you can connect to one of the workers and run \texttt{sudo tmux a -t install}, and sure that the pane has completed all of its commands.
Alternatively, you can wait around 20 minutes, and this will most likely be long enough for the image to complete building.

\subsubsection{Orchestrator}
Building the orchestrator requires slightly more interaction from the user.

\texttt{bash orchestrator\_deploy.sh <orchestrator\_host> 22}

First, like before we will set up the environment using a deployment script. This will, again, automate the file transfer and environment setup.

Next, connect to your orchestrator node and run the following commands.
Note, that the first command is \texttt{tmux}.
We recommend using this as tuning runs are long-running.
Without \texttt{tmux}, disconnects over \texttt{ssh} are possible, however this is not technically required.\\
\texttt{
tmux\\
make -C src/MLOS conda-env\\
conda activate mlos
}

\subsection{Evaluation workflow}
\subsubsection{Major Claims}
\textit{We list the major claims and corresponding experiments below:}\\

\begin{itemize}
    \item \textit{(C1)}: sampling variability significantly decreases the average rate of convergence. This is proven by the experiment (E1) described in \ref{sec:cloud_convergence}.
    \item \textit{(C2)}: \SystemName can generalize across various workloads and achieve better performance, lower variability, or both compared to traditional sampling and the default configuration.
    This is proven by the experiment (E2) shown in Figure~\ref{fig:eval_workloads}.
    \item \textit{(C3)}: \SystemName can generalize across various sets of execution environments and achieve better performance, lower variability, or both compared to traditional sampling and default configuration.
    This is proven by the experiment (E3) shown in Figure~\ref{fig:eval_cloudlab} and Figure~\ref{fig:eval_cus}.
    \item \textit{(C4)}: \SystemName can generalize across various systems and achieve better performance, lower variability, or both compared to traditional sampling and default configuration.
    This is proven by the experiment (E4) shown in Figure~\ref{fig:eval_redis} and Figure~\ref{fig:eval_nginx}.
    \item \textit{(C5)}: \SystemName can generalize across different optimizers and achieve better performance, lower variability, or both compared to traditional sampling and default configuration.
    This is proven by the experiment (E5) shown in Figure~\ref{fig:eval_gp}.
\end{itemize}

\subsubsection{Experiments}
~\\
\textit{Experiment (E1): [5 human-minutes + 800 compute hours]: This experiment compares the rate of convergence for tuning epinions with various levels of synthetic sampling noise.}\\
\textit{[Preparation]}
\textit{
Set up the environment as described above.}\\
\textit{[Execution]}
\textit{Run the \texttt{parallel\_prior.py} command described in the \texttt{README} for three levels of noise: $0\%$, $5\%$ and $10\%$}.\\
\textit{[Results]}
\textit{The results should show that as the noise levels increase, the rate of convergence decreases.}\\

~\\
\textit{Experiment (E2): [20 human-minutes + 4000 compute hours]: This experiment compares the average performance and average standard deviation of various configurations found for various workloads on \postgres.}\\
\textit{[Preparation]}
\textit{
Set up the environment as described above.}\\
\textit{[Execution]}
\textit{Run the \texttt{parallel.py} and \texttt{python3 TUNA.py} command as described in the \texttt{README} }. This will collect the results into data frames. These results can then be aggregated, and deployment can be run on, ideally, a new cluster using \texttt{mass\_reruns\_v2.py} \\
\textit{[Results]}
\textit{The results should show \SystemName improves the target performance metrics, decreases variability, or both.}\\

~\\
\textit{Experiment (E3): [20 human-minutes + 2000 compute hours]: This experiment compares the average performance and average standard deviation of various configurations found for TPC-C on \postgres running on CloudLab Nodes and in a new Azure region.}\\
\textit{[Preparation]}
\textit{
Set up the environment as per the above instructions in both a new Azure region and CloudLab.}\\
\textit{[Execution]}
\textit{Run the \texttt{parallel.py} and \texttt{python3 TUNA.py} command as described in the \texttt{README} } in both environments. This will collect the results into data frames. These results can then be aggregated, and deployment can be run on, ideally, a new cluster using \texttt{mass\_reruns\_v2.py} \\
\textit{[Results]}
\textit{The results should show \SystemName improves the target performance metrics and decreases variability on CloudLab Nodes and in a new Azure region.}\\

~\\
\textit{Experiment (E4): [20 human-minutes + 2000 compute hours]: This experiment compares the average performance and average standard deviation of various selected configurations found \redis and \nginx.}\\
\textit{[Preparation]}
\textit{
Set up the environment as described above.}\\
\textit{[Execution]}
\textit{Run the \texttt{parallel.py} and \texttt{python3 TUNA.py} command as described in the \texttt{README} } for both systems. This will collect the results into data frames. These results can then be aggregated, and deployment can be run on, ideally, a new cluster using \texttt{mass\_reruns\_v2.py} \\
\textit{[Results]}
\textit{The results should show \SystemName improves the target performance metrics and decreases variability on CloudLab Nodes and in a new Azure region.}\\

~\\
\textit{Experiment (E5): [20 human-minutes + 1000 compute hours]: This experiment compares the average performance and average standard deviation of various configurations found for various workloads on \postgres using a Gaussian process optimizer.}\\
\textit{[Preparation]}
\textit{
Set up the environment as described above.}\\
\textit{[Execution]}
\textit{Run the \texttt{paralle\_gp.py} and \texttt{python3 TUNA\_gp.py} command as described in the \texttt{README} }. This will collect the results into data frames. These results can then be aggregated, and deployment can be run on, ideally, a new cluster using \texttt{mass\_reruns\_v2.py} \\
\textit{[Results]}
\textit{The results should show \SystemName improves the target performance metrics, decreases variability, or both using a Gaussian Process optimizer.}\\

\subsection{Notes on Reusability}
\label{sec:reuse}
If you want to further extend these to new systems, they simply need to be added to the version of Nautilus included in the artifact.
A good place to start is to compare the \redis implementation with the \postgres implementation to understand how different systems interact with Nautilus, and how the new target system could be added.
In the future we intend to integrate \SystemName to MLOS~\footnote{\url{https://github.com/microsoft/MLOS/issues/926}} as well.
\end{appendices}

\end{document}